\documentclass[12pt]{article}
\usepackage{jheppub}
\usepackage{iip-arxiv}
\usepackage[utf8]{inputenc}
\usepackage{bm}
\usepackage{mathtools}
\usepackage{physics}
\usepackage{xparse}  
\usepackage{stackengine}

\makeatletter
\pdfoutput=1

\def\subrangle#1{\stackengine{5pt}{}{$\!\scriptstyle #1$}{U}{l}{F}{F}{L}}
\NewDocumentCommand{\prim}{O{v} O{P,a} m}{%
|#1^{\phantom{*}}_{#2}\rangle
\IfValueT{#3}{\subrangle{#3}
}%
}
\NewDocumentCommand{\dualprim}{O{w^{*}} O{P,a} m}{%
  \IfValueT{#3}{\subrangle{#3}}\langle #1_{#2}|}
\NewDocumentCommand{\inner}{m o m o}{%
\IfValueT{#2}{\subrangle{#2}}%
\mspace{-1.5mu}\langle #1\!\mid\! #3\rangle
\IfValueT{#4}{\subrangle{#4}}%
}
\newcommand{\del}[1][z]{\partial_{#1}}
\newcommand{\hilbert}[1][u]{\mathcal{D}^{H}_{#1}}
\newcommand{\pvint}{\mathop{\mathcal{P}}\!\int}
\newcommand{\cauchy}[2][z]{\oint_{#2}\frac{d#1}{2\pi i}\;}

\title{Generalized Gibbs Ensemble of 2D CFTs with U(1) Charge
  from the AGT Correspondence}

\author[1]{Fábio Novaes}
\affiliation{Centro de Estudios Científicos, Valdivia, Chile}
\emailAdd{novaes at cecs.cl}

\abstract{The Generalized Gibbs Ensemble (GGE) is relevant to
  understand the thermalization of quantum systems with an infinite
  set of conserved charges. In this work, we analyze the GGE partition
  function of 2D Conformal Field Theories (CFTs) with a U(1) charge
  and quantum Benjamin-Ono$_{2}$ (qBO$_{2}$) hierarchy charges. We use the
  Alday-Gaiotto-Tachikawa (AGT) correspondence to express the thermal
  trace in terms of the Alba-Fateev-Litvinov-Tarnopolskiy (AFLT) basis
  of descendants, which diagonalizes all charges.  We analyze the GGE
  partition function in the thermodynamic semiclassical limit,
  including the first order quantum correction. We find that the
  equality between GGE averages and primary eigenvalues of the
  qBO$_{2}$ charges is attainable in the strict large $c$ limit and
  potentially violated at the subleading $1/c$ order. We also obtain
  the finite $c$ partition function when only the first non-trivial
  charge is turned on, expressed in terms of partial theta functions.
  Our results should be relevant to the eigenstate thermalization
  hypothesis for charged CFTs, Warped CFTs and effective field theory
  descriptions of condensed matter systems.}

\preprint{CECS-PHY-21/01}

\makeatother

\begin{document} 
\maketitle

\baselineskip=17pt

\section{Introduction}
\label{sec:introduction}

The emergence of thermodynamics from interacting quantum systems is
one of the most fundamental and intriguing topics in physics. While
open quantum systems can be argued to equilibrate with an external
heat bath, isolated quantum systems are harder to adjust to the usual
statistical mechanics arguments.
The traditional understanding is that quantum integrable systems do
not thermalize due to the existence of a infinite number of conserved
charges, contrary to chaotic systems. This notion can be put into more
concrete terms in a precise hypothesis: if an isolated quantum system
obeys the eigenstate thermalization hypothesis (ETH)
\cite{deutsch1991quantum,srednicki1994chaos,Deutsch2018}, it should
dynamically thermalize. The ETH is essentially a statement on the
condition for quantum expectation values to approximately match
ensemble averages in the thermodynamic limit. More precisely, given a
local operator $\mathcal{O}$ and an energy eigenbasis $|E_{n}\rangle$,
the ETH can be mathematically stated as
\begin{equation}
  \label{eq:eth-condition}
  \langle E_{n}|\mathcal{O}|E_{m}\rangle = \mathcal{O}_{\text{th}}\delta_{nm}
  + e^{-S}\delta_{\mathcal{O}}(E_{n},E_{m})R_{nm},
\end{equation}
where $\mathcal{O}_{\text{th}}$ is a thermal average, $S$ is the
entropy, $\delta_{\mathcal{O}}$ is a slowly varying function of its
arguments and $R_{nm}$ is a random matrix of zero mean and unit
variance \cite{Deutsch2018}.  Strong universal properties are supposed
to exist to make this hypothesis true for a large class of quantum
systems. In the case the theory has global symmetries, we should
consider the charged version of ETH. For neutral operators, the
charged ETH is essentially equivalent to the version above with the
appropriate dependence on the charges
\cite{Belin:2020jxr}.

In the context of 2D conformal field theories (CFTs), one can argue
that minimal models are integrable and do not thermalize
\cite{de2016remarks}. On the other hand, given the AdS/CFT duality,
one expects that CFTs with large central charge should thermalize,
consistent with black holes behaving as thermal systems.  If one
restricts the attention to the operators in the CFT vacuum family,
detailed investigations showed that the ETH is indeed satisfied in the
large c limit of holographic 2D CFTs
\cite{Lashkari:2016vgj,Basu_2017,Lashkari_2018}. This match claims
that such CFTs are chaotic
\cite{Roberts:2014ifa,Maldacena:2015waa,Turiaci:2016cvo}, suggesting
distinctive criteria for a CFT to have a gravitational dual. The
justification for focusing on the vacuum family is that the thermal
expectation values for generic operators should vanish in the
thermodynamic limit. However, there are indications that the ETH
matching for the vacuum family fails for finite central charge in the
usual micro and canonical ensembles
\cite{Lashkari:2016vgj,Basu_2017,Lashkari_2018}. This opens up the
question of what is the proper formulation of ETH for 2D CFTs.
    
For generic primary operators O, the matrix elements in (1.1) are
related to OPE coefficients. Thus, the most general formulation of ETH
for 2D CFTs is one about the statistics of OPE coefficients. One can
then prove an averaged version of ETH by integrating these
coefficients over an order one energy window
\cite{Collier:2019weq,Das:2020uax}. However, the original ETH equality
between ensembles has to be proven in a exponentially small energy
window, i.e., effectively for a single eigenstate, and this has not
been proven in full generality yet (see \cite{Garrison:2015lva} for a
related discussion outside the CFT context)\footnote{We thank the
  anonymous referee for these remarks on the ETH for generic
  operators.}. In this work, we focus on questions of typicality of
states in a thermal ensemble with conserved charges, which might be
useful to analyze the ETH for the vacuum family, but not necessarily
for generic operators.
 
2D Virasoro CFTs have an infinite set of commuting charges from the
quantum Korteweg-de Vries (qKdV) hierarchy, evidence of a rich
integrable structure
\cite{Bazhanov:1994ft,bazhanov1997integrable,bazhanov1999integrable}.
That could be the reason why the ensemble matching mentioned above
fails for finite $c$, as a large class of states keep memory of its
initial conditions via those integrable charges \cite{Cardy_2016}. The
notion of a generalized ETH has been proposed to account for
equilibration in quantum integrable systems
\cite{Rigol:2007aa,rigol2008thermalization,cassidy2011generalized,Vidmar_2016,D_Alessio_2016}. This
formulation compares eigenstate averages with Generalized Gibbs
Ensemble (GGE) averages, instead of the canonical ensemble, which is
defined as a grand canonical ensemble with an infinite set of charges
and associated chemical potentials.

In this work, we will consider the GGE defined in terms of mutually
commuting charges $\{Q_{k}\}_{k\geq 0}$ obtained from  symmetry
currents, where $Q_{0}\equiv \mathcal{J}$ is a U(1) current,
$Q_{1} \equiv H$ is the CFT Hamiltonian and $\{Q_{k}\}_{k>1}$ form an
infinite tower of higher spin conserved charges. We want to compare
the averages over some common eigenstate $|E,q \rangle$ of
all charges
\begin{equation}
  \label{eq:eth-average}
  \langle E,q |\mathcal{O}|E,q\rangle \equiv f_{\mathcal{O}}(Q_{k}(E,q)), 
\end{equation}
where $E$ is the energy, $q$ is the U(1) charge and $f_{\mathcal{O}}$
is a smooth function of the charges, with a generalized Gibbs ensemble
average
\begin{equation}
  \label{eq:gge-average}
  \langle \mathcal{O}\rangle _{\text{GGE}}
  \equiv
  \frac{1}{Z(\bm{\eta})}\Tr
  \left(
    \mathcal{O}e^{\sum_{k\geq 0}\eta_{k}Q_{k}}
  \right), 
\end{equation}
where $\bm{\eta} = (\eta_{0},\eta_{1},\eta_{2},\dots)$ are generalized
chemical potentials, with $\eta_{0} \equiv \eta$ being the U(1)
chemical potential, $\eta_{1} \equiv -\beta$ being the inverse
temperature and the partition function given by
\begin{equation}
  \label{eq:partition-function}
  Z(\bm{\eta})=\Tr
  \left(
   e^{\sum_{k\geq 0}\eta_{k}Q_{k}}
  \right). 
\end{equation}
In particular, we will focus on the averages of the
conserved charges themselves, i.e. $\mathcal{O} = Q_{k}$, such that
\begin{equation}
  \label{eq:12}
  \langle Q_{k}\rangle(\bm{\eta}) = \del[\eta_k]\log  Z(\bm{\eta}).
\end{equation}
The qKdV GGE and its generalized ETH have been recently investigated
in the 2D CFT literature by
\cite{Cardy_2016,de2016remarks,Dymarsky:2018lhf,Maloney:2018yrz,
  Maloney:2018hdg,Dymarsky:2018iwx,Brehm:2019fyy,Dymarsky:2019etq}.
The overall conclusion is that the generalized ETH also only works at
leading order in the large central charge limit, in particular for the
vacuum family, i.e., descendants of the identity operator. Therefore,
our results should be interesting to analyze these quasi-primary
operators, but do not address the stronger statements on the
generalized ETH for generic operators, which still remains as an open
problem. We should also mention the papers
\cite{Datta:2019jeo,Besken:2019bsu}, which discuss CFT thermalization
arguing that heavy descendants are the most typical states in the
thermodynamic limit.  Here we will consider that the typical
eigenstates in the thermodynamic semiclassical limit are primaries.

A crucial challenge in the investigation of generalized ETH for 2D
CFTs is having better control of finite $c$ GGE averages. We provide
advancements in this direction focusing on 2D CFTs with a U(1)
current, henceforth called charged CFTs. A natural integrable
hierarchy associated to this case is the quantum Benjamin-Ono$_{2}$
(qBO$_{2}$) hierarchy \cite{Abanov:2008ft,Alba:2010qc}, whose
relevance came to light via a CFT proof \cite{Alba:2010qc} of the AGT
correspondence \cite{Alday:2009aq}.  This correspondence states the
equivalence of partition functions of $\mathcal{N}=2$ four-dimensional
supersymmetric quiver gauge theories and Liouville correlation
functions, using Nekrasov's instanton partition functions
\cite{Nekrasov:2002qd}.  The full AGT correspondence reproduces
Liouville theory correlators, including its precise structure
constants. On the other hand, it also provides an alternative
combinatorial expansion for CFT conformal blocks using the
Alba-Fateev-Litvinov-Tarnopolskiy (AFLT) basis of
Vir$\times$$\mathfrak{u}$(1) descendants \cite{Alba:2010qc}, where Vir
is the Virasoro algebra. Perhaps surprisingly, the AFLT basis
diagonalizes all the
qBO$_{2}$ conserved charges. The complete set of eigenvalues of this
hierarchy has been conjectured in \cite{Litvinov:2013zda} as a
particular case of the quantum
$\mathfrak{gl}$(2) Intermediate Long Wave
(qILW$_{2}$) hierarchy. The
qILW$_{2}$ spectrum was defined in terms of Bethe ansatz equations
later proven in \cite{Feigin:2017gcv} (see also
\cite{Litvinov:2020zeq}). The AFLT basis allow us to obtain exact
expressions for matrix elements and the GGE partition function at
finite
$c$, which is the major technical point of this paper. We notice that
the charges obtained in \cite{Alba:2010qc} did not include the U(1)
zero modes, relevant for our GGE discussion, and so we fill this gap
by deriving the first three
qBO$_{2}$ charges with U(1) zero modes in appendix
\ref{sec:deriv-integr-moti}. To achieve this, we slightly reformulate
the algorithm of \cite{Dymarsky:2019iny} in terms of OPEs and
including the U(1) terms.

We also discuss the validity of the generalized ETH for the
qBO$_{2}$ charges in the sense of \eqref{eq:eth-condition}. This
requires analyzing the thermodynamic and large central charge limits
of the conserved charges. Using a saddle-point calculation, we reach
the conclusion that the generalized ETH for the charges is valid in
the strict large central charge limit, similar to the qKdV results
\cite{Dymarsky:2018lhf,Maloney:2018yrz}. In particular, if we discard
the U(1) zero modes, the
qBO$_{2}$ eigenvalues reduce to the qKdV ones in the thermodynamic
semiclassical limit, and thus both ensembles become equivalent (see
sec.~\ref{sec:saddle-point-semicl}).  The sub-leading corrections in
the central charge can also be written in terms of free boson
representations, similar to the qKdV case \cite{Dymarsky:2018iwx}.
However, here we see that the finite $c$
qBO$_{2}$ eigenvalues have more complicated free boson
representations, briefly discussed in appendix~\ref{sec:comb-ident},
which complicates the full integration of the partition function. We
identify a potential violation of the generalized ETH for the first
$1/c$ correction, but we had to rely on a particular ansatz for the
saddle-point, leaving this question still open in general. Therefore,
this work sets a new route to investigate the generalized ETH for
charged CFTs at finite
$c$ using the AGT correspondence. Other potential applications of our
results are on generalized thermalization of Warped CFTs
\cite{Detournay:2012pc} and holographic realizations of integrable
hierarchies
\cite{Perez:2016vqo,Fuentealba:2017omf,Melnikov:2018fhb,Ojeda:2019xih,Dymarsky:2020tjh}. The
quantum BO$_{2}$ hierarchy has also been conjectured to describe non-linear
dynamics of quantum liquids
\cite{abanov2005quantum,bettelheim2006quantum,bettelheim2007nonlinear,Abanov:2008ft,wiegmann2012nonlinear}
and we hope our results can also be relevant in this context.

Here we review the contents of this paper. In section
\ref{sec:bo2-hierarchy}, we review the qBO$_{2}$ hierarchy, its
conserved charges, eigenstates and eigenvalues. We use the Liouville
parametrization for the CFT parameters, as done in the AGT
correspondence, distinguishing the two unitary branches of the
spectrum. We show explicitly the first three qBO$_{2}$ eigenvalues in
terms of the Liouville momentum, the U(1) charge and the two
partitions labelling descendant states. These eigenvalues are
generically complex and we discuss conditions to make them real, as
expected for equilibrium states.  In section
\ref{sec:gener-gibbs-ensemble}, we analyze the qBO$_{2}$ GGE in the
thermodynamic limit. We remark that the eigenvalues have different
limits depending on the unitary branch of the spectrum. We then obtain
the free energy in the large central charge limit and its first $1/c$
correction from a saddle-point calculation. We obtain explicit
expressions assuming that only the first two non-trivial chemical
potentials $\eta_{2}$ and $\eta_{3}$ are non-zero. We also obtain an
exact finite $c$ partition function when only $\eta_{2}$ is non-zero,
written in terms of partial theta functions after integrating over the
descendants. In section \ref{sec:geth-qbo_2-charges}, we discuss the
generalized ETH for the qBO$_{2}$ charges. We reach the conclusion
that it is valid in the strict large central limit, but that it
potentially fails at the $1/c$ order. Finally, in section
\ref{sec:conclusions}, we discuss how our results can be used to study
generalized thermalization of 2D charged CFTs, Warped CFTs and
effective field theories in condensed matter systems. In appendix
\ref{sec:deriv-integr-moti}, we review how to derive the qBO$_{2}$
conserved charges using analytic ordering and in appendix
\ref{sec:comb-ident}, we derive some useful combinatorial identities
for the analysis of the descendant eigenvalues.

\section{The Quantum Benjamin-Ono$_{2}$ Integrable Hierarchy}
\label{sec:bo2-hierarchy}

The notion of integrability has many forms in quantum field theories
(QFTs). In its simplest definition, the integrable structure appears
as an infinite set of mutually commuting conserved charges built out
of a finite number of fundamental fields. The KdV equation is part of
an integrable hierarchy, an infinite set of equations of motion
generated by the aforementioned set of conserved charges. While the
well-known KdV equation describes non-linear waves in a shallow
channel, the not so widely known Benjamin-Ono (BO) equation describes
internal waves in a deep channel
\cite{benjamin1967internal,ono1975algebraic}.  The Intermediate Long
Wave (ILW) equation \cite{joseph1978multi,Chen:1979aa} is a
generalization of these two equations interpolating between both
regimes, with the channel depth as a free parameter. The ILW equation
is also part of an integrable hierarchy generalizing the BO and KdV
ones, which in its turn can be  generalized to hierarchies with a
multiple number of fundamental fields \cite{Lebedev:1982rj}.  A
generalization of the BO hierarchy with two fundamental fields has
been rediscovered from the analysis of Calogero-Sutherland fluids in
\cite{Abanov:2008ft}, henceforth called the BO$_{2}$ hierarchy,
proposed to be relevant for effective low-energy descriptions of
condensed matter systems.

All the above mentioned classical integrable hierarchies have a
quantum version, which can be built out of Virasoro and current
algebras. In the case of 2D CFTs, the natural integrable hierarchy is
the qKdV hierarchy
\cite{Bazhanov:1994ft,bazhanov1997integrable,bazhanov1999integrable},
whose conserved charges are integrals of appropriate polynomials of
the energy-momentum tensor $T(z)$ and its derivatives (similarly to
its anti-holomorphic counterparts). Likewise, the qBO$_{2}$ integrable
hierarchy is a natural one for 2D CFTs with an extra U(1) current
$J(z)$ \cite{Alba:2010qc}. As in the classical case, the
 qILW$_{2}$ hierarchy gives the qBO$_{2}$ hierarchy in
one limit and the commuting qKdV and free boson hierarchies in the
other limit. In this work, we focus on the qBO$_{2}$ hierarchy as its
spectrum has a simpler form than the qILW$_{2}$ one, both derived in
\cite{Litvinov:2013zda}. For other interesting applications of these
integrable hierarchies in supersymmetric theories, see, for example,
\cite{Bonelli:2014iza,Koroteev:2015dja,Koroteev:2016znb,Gorsky:2019yqp}.

The qBO$_{2}$ charges were first calculated in \cite{Alba:2010qc} and
its eigenvalues were obtained in \cite{Litvinov:2013zda}. Let us
review this construction in the following, with the particular
addition of the U(1) zero modes not considered before.

\subsection{AFLT basis for Descendants}
\label{sec:new-basis-desc}

The tensor product algebra
$\mathcal{A} = \text{Vir}\otimes \mathfrak{u}(1)$, where Vir is the
Virasoro algebra, is generated by the Virasoro and U(1) currents
defined on the cylinder
\begin{equation}
  \label{eq:1}
  T(x) = \sum_{n=-\infty}^{\infty}L_{n}e^{-inx} - \frac{c}{24},\quad
  J(x) = \sum_{n=-\infty}^{\infty}a_{n}e^{-inx},\quad x\in (0,2\pi),
\end{equation}
with modes obeying the commutation relations 
\begin{equation}
  \label{eq:AFLTalgebra}
  \begin{aligned}
    [L_{n},L_{m}] &= (n-m)L_{n+m} +
    \frac{c}{12}n(n^{2}-1)\delta_{n,-m},\\[5pt]
    [a_{n},a_{m}]&=\frac{n}{2}\delta_{n,-m},\quad [a_{n},L_{m}]=0.
  \end{aligned}
\end{equation}
A primary state is here denoted by $\prim{}$ and obeys the relations
\begin{equation}
  \label{eq:11}
  \begin{aligned}
    a_{0}\prim{} &= a\prim{},&
    a_{n}\prim{} &=0,\; n > 0,\\[5pt]
    L_{0}\prim{} &= \Delta(P) \prim{},&
    L_{n} \prim{} &= 0,\; n > 0, 
  \end{aligned}
\end{equation}
where the Liouville parametrization for the central charge $c$ and
conformal weights $\Delta$ is given by
\begin{equation}
  \label{eq:liouville-notation}
  c=1+6Q^{2},\quad  Q= b+\frac{1}{b},\quad \Delta(P) = \frac{Q^{2}}{4}-P^{2},
\end{equation}
with Liouville parameter $b$ labeling the central charge and the
momentum $P$ labeling the conformal weights $\Delta(P)$ of primary
states. An equivalent definition follows for the anti-holomorphic
modes with eigenvalues $\bar{\Delta}(\bar{P})$ and $\bar{a}$. These
conventions can be used to describe the spectrum of any CFT if we
properly specify $b$ and the domain of $P$ and $a$, which can also be
discrete in the case of minimal models (for a recent CFT treatise
using this notation, see \cite{Ribault:2014hia}). We also mention that
the Liouville parametrization \eqref{eq:liouville-notation} is
invariant under $b\rightarrow 1/b$. Thus, there are two ways to get
the semiclassical limit $c\rightarrow \infty$, using either
$b\rightarrow 0$ or $b\rightarrow \infty$. In this work, we will only
consider the $b\rightarrow 0$ limit.

Assuming that $c>1$, we distinguish two unitary branches of the
spectrum with $\Delta > 0$
\begin{equation}
  \label{eq:spectrum-branches}
  \begin{aligned}
    \text{Real branch:}\quad \mathcal{S}_{R} &= \{ P\in
    \mathcal{S}\subseteq
    \left(-\frac{Q}{2},\frac{Q}{2}\right),\; Q \in \mathbb{R}\},\\[5pt]
    \text{Liouville branch:}\quad \mathcal{S}_{L}&= \{ P = ip,\quad p
    \in\mathcal{S} \subseteq\mathbb{R}\},
  \end{aligned}
\end{equation}
where $\mathcal{S}$ is some discrete or continuum subset of
$\mathbb{R}$.  The AGT formulas are more conveniently written with
respect to $P$, but later we will focus on the unitary thermodynamic
semiclassical limit, which is only attainable in the Liouville branch
$P=ip$. To avoid extra degeneracy in the spectrum, we shall consider
$p \geq 0$. With respect to the U(1) charge $a$, we shall initially
assume that it is either real or purely imaginary. As we will see
below, the spectrum of the qBO$_{2}$ charges can be complex and
unbounded. The mitigation of these potentially unphysical properties
depends on further assumptions on the domains of $p$ and $a$, related
to the convergence of the partition function.  We do not have a
general solution for this, but we will exemplify how it can be solved
for the first two non-trivial qBO$_{2}$ charges in section
\ref{sec:herm-equil}. We will also discuss convergence properties of
the partition function and boundedness of the free energy in section
\ref{sec:gener-gibbs-ensemble}.

The standard basis for the descendant states, henceforth called just
descendants, is
\begin{equation}
  \label{eq:virasoro-descendants}
 a_{-\lambda_{m}}\cdots a_{-\lambda_{1}}L_{-\mu_{n}}\cdots L_{-\mu_{1}}
  \prim{},\quad
 \lambda_{1}\geq \lambda_{2}\geq
 \cdots \geq  \lambda_{m}>0,\quad \mu_{1}\geq \mu_{2} \geq \cdots \geq \mu_{n}>0,\quad
\end{equation}
where $\lambda = (\lambda_{1},\lambda_{2},\dots,\lambda_{m})$ and
$\mu = (\mu_{1},\mu_{2},\dots,\mu_{n})$ are two independent partitions
of an integer. To simplify the notation, we introduce a multi-index
$I = (\lambda,\mu)$ such that a descendant at level
$|I| \equiv |\lambda| + |\mu|$, with $|\lambda| = \sum_{i}\lambda_{i}$
and $|\mu| = \sum_{i}\mu_{i}$, is defined by
\begin{equation}
  \label{eq:descendant-basis}
  \prim{I} \equiv
  a_{-\lambda_{m}}\cdots a_{-\lambda_{1}}L_{-\mu_{n}}\cdots L_{-\mu_{1}}
  \prim{}.
\end{equation}
This basis has a natural Hermitean inner product if we use the
prescription 
\begin{equation}
  \label{eq:hermiticity-condition}
  L_{n}^{\dagger} = L_{-n},\quad a_{n}^{\dagger} = a_{-n},
\end{equation}
such that the dual vector of \eqref{eq:descendant-basis} is
\begin{equation}
  \label{eq:83}
  \dualprim[v]{I} \equiv \dualprim[v]{}
  L_{\mu_{n}}\cdots L_{\mu_{1}}a_{\lambda_{m}}\cdots a_{\lambda_{1}}.
\end{equation}
The Kac-Shapovalov matrix of inner products is then level-orthogonal
\begin{equation}
  \label{eq:81}
  M_{I,J}(\Delta(P)) = \inner{v_{P,a}\!}[I]{\!v_{P,a}}[J]
  \propto \delta_{|I|,|J|}\,,
\end{equation}
where the Kronecker delta denotes its block-diagonal form. As the
Virasoro and U(1) algebras commute, the inner product decomposes into
two independent inner products. The U(1) inner product is positive
definite in our conventions and the U(1) zero mode plays no role in
this case. The Virasoro inner product is also positive definite for
$c>1$ if $\Delta>0$ \cite{Belavin:1984vu,DiFrancesco1997a}. Therefore,
there are no null or negative states if $c>1$, $\Delta>0$ and $a$ arbitrary,
implying that the CFT spectrum is unitary.

The usual convention in the charged CFT literature is to define the
Virasoro modes in terms of a total energy-momentum tensor
$T' = T + \frac{J^{2}}{\kappa}$, with $\kappa$ being the Kac-Moody
level. In this case, the Virasoro modes do not commute with the U(1)
modes
\begin{equation}
\label{eq:virasoro-kac-moody-algebra}
\begin{aligned}
  [L'_{n},L'_{m}] &= (n-m)L'_{n+m} +
  \frac{c+1}{12}n(n^{2}-1)\delta_{n,-m},\\[5pt]
  [a_{n},a_{m}]&=\kappa\frac{n}{2}\delta_{n,-m},\quad [a_{n},L'_{m}]=n
  a_{n+m},
\end{aligned}
\end{equation}
and the spectrum unitarity depends on the domains of both $P$ and $a$,
which is particularly relevant to the study of Warped CFTs
\cite{Detournay:2012pc,Castro:2015uaa,Apolo:2018eky}. However, as we
discuss in section \ref{sec:gener-gibbs-ensemble}, the partition
function is invariant under this change of basis and our results are
also valid for \eqref{eq:virasoro-kac-moody-algebra}. Therefore, we
will mostly focus on the commuting algebra \eqref{eq:AFLTalgebra},
equivalent to \eqref{eq:virasoro-kac-moody-algebra} with $\kappa=1$ up
to this change of basis, which greatly simplify the computations.

Now, we introduce the AFLT basis \cite{Alba:2010qc} as
special linear combinations of the descendants
\eqref{eq:descendant-basis}
\begin{equation}
  \label{eq:aflt-basis}
  \prim[w]{I}=
  \sum_{\substack{J\in\mathbb{Y}_{|I|}}}C_{I}^{J}(P)\prim{J}\,,
\end{equation}
where $\mathbb{Y}_{|I|}$ is the set of all partitions of size $|I|$.
The first few coefficients $C_{I}^{J}(P)$ are given in
\cite{Alba:2010qc} and can be derived from the recursive algorithm
described there.  The Hermitean conjugate of \eqref{eq:aflt-basis}
depends on the proper domain of $P$. However, in the AFLT
construction, the dual basis is defined without complex conjugation of
the coefficients as
\begin{equation}
  \label{eq:dual-aflt-basis}
  \dualprim{I}=
  \sum_{\substack{J\in\mathbb{Y}_{|I|}}}\dualprim[v]{J} C_{I}^{J}(P)\,,
\end{equation}
which we distinguish  from the usual Hilbert space dual basis
${}_{I}\langle w_{P,a}|$ by a star in the vector.

Given a primary field $V_{\alpha}$, the crucial fact of the AFLT basis
is that
\begin{equation}
  \label{eq:aflt-theorem}
\frac{ \dualprim[w][P',a]{I} V_{\alpha}  \prim[w]{J}
}{\dualprim[w][P',a]{} V_{\alpha} \prim[w]{} } =
Z_{\text{bif}}(\alpha | P',I; P, J ),
\end{equation}
where $Z_{\text{bif}}$ is the bifundamental part of the Nekrasov
partition function~\cite{Nekrasov:2002qd} defined as
\begin{equation}
  \label{eq:nekrasov-bif}
  Z_{\text{bif}} (\alpha | P', I; P, J)
  = \prod_{i,j=1}^{2}\prod_{s\in \mu_{i}}[Q -
  E_{\mu_{i},\lambda_{j}}(\sigma_{i}P-\sigma_{j}P'|s)-\alpha]
  \prod_{t\in \lambda_{j}}[ E_{\lambda_{j}, \mu_{i}}(\sigma_{j}P'-\sigma_{i}P|t)-\alpha]
\end{equation}
where $I =(\lambda_{1},\lambda_{2})$, $J=(\mu_{1},\mu_{2})$,
$(\sigma_{1},\sigma_{2})=(1,-1)$ and
\begin{equation}
  \label{eq:E-function}
  E_{\lambda,\mu}(P|s) = P- b l_{\mu}(s) + b^{-1}(a_{\lambda}(s)+1).
\end{equation}
For a given a box $s=(i,j)$, $a_{\lambda}(s) = \lambda_{i}-j$ refers
to the arm length of the box $s$ in the partition $\lambda$ and
$l_{\mu}(i,j)=\mu'_{j}-i$ is the leg length of the box $s$ in the
partition $\mu$, where $\mu'$ is the transpose partition of $\mu$. One
can also show that \cite{Alba:2010qc}
\begin{equation}
  \label{eq:aflt-orthogonality}
\inner{w^{*}_{P',a}}[I]{w^{\phantom{*}}_{P,a}}[J]
  =\mathcal{N}_{I}(P)\,
  \delta_{P',P}\,\delta_{I,J},\quad
  \mathcal{N}_{I}(P) \equiv Z_{\text{bif}}(0 | P, I; P, I),
\end{equation}
which means that the AFLT basis is orthogonal in this inner
product. As a consequence of \eqref{eq:aflt-basis},
\eqref{eq:dual-aflt-basis} and \eqref{eq:aflt-orthogonality}, we have
\begin{align}
   \label{eq:84}
  \sum_{K,L} C^{K}_{I}(P)M_{K,L}(\Delta(P))C^{L}_{J}(P) = \mathcal{N}_{I}(P)\delta_{I,J},
\end{align}
or in matrix form
\begin{align}
  \label{eq:diagonalization-kac-matrix}
  C\cdot M\cdot C = \mathcal{N},
\end{align}
where $\mathcal{N}$ is diagonal, 
making explicit that the transformation \eqref{eq:aflt-basis}
diagonalizes the inner product matrix, although not in the usual
Hermitean conjugate dual basis.

Taking
the determinant of \eqref{eq:diagonalization-kac-matrix}, we notice
that $\mathcal{N} = 0$ only for degenerate states \cite{Alba:2010qc}
and those are out of the unitary spectrum for $c>1$. Therefore, we can
redefine
$\dualprim{I} \rightarrow \sqrt{\mathcal{N}_{I}(P)}\;\dualprim{I}$ and
make the basis orthonormal. The AFLT states with
$I = (\lambda_{1},\varnothing)$ and $I = (\varnothing, \lambda_{2})$
are equivalent to the Jack basis, eigenfunctions of the
Calogero-Sutherland (CS) Hamiltonians \cite{Sakamoto:2004rn}. The
complete basis has an interpretation in terms of eigenstates of two
interacting CS hamiltonians, obtained using a bosonic representation of the
Virasoro generators \cite{estienne2012conformal}.

\subsection{qBO$_2$ Charges and  Eigenvalues}
\label{sec:qbo_2-charg-eigenv}

The AFLT states \eqref{eq:aflt-basis} are eigenstates of the qBO$_{2}$
Hamiltonians $Q_{k}$ \cite{Alba:2010qc}
\begin{equation}
  \label{eq:4}
  Q_{k} = \frac{1}{2\pi}\int_{0}^{2\pi}dx \,\mathcal{Q}_{k+1}(x),\quad
  k=0,1,2,\dots,
\end{equation}
obtained by integrating the currents $\mathcal{Q}_{k}(x)$, with the first
four ones  given by
\begin{equation}\label{Quantum-densities}
  \begin{aligned}
    &\mathcal{Q}_{1}=J,\\
    &\mathcal{Q}_{2}=T+J^{2},\\
    &\mathcal{Q}_{3}=TJ+ i
    Q J\mathcal{H}J_{x}+\frac{1}{3}J^{3},\\
    &\mathcal{Q}_{4}=T^{2}+6TJ^{2}+6iQT\mathcal{H}J_{x}+6iQ
    J^{2}\mathcal{H}J_{x}-6Q^{2}(\mathcal{H}J_{x})^{2}+(1+Q^{2})J_{x}^{2}+J^{4},
  \end{aligned}
\end{equation}
where $J_{x} \equiv \frac{dJ}{dx}$ and $\mathcal{H}$ represents the
Hilbert transform on the circle 
\begin{equation}
  \label{eq:hilbert-transform-circle}
  \mathcal{H}f(x) = \frac{1}{2\pi} \pvint_{0}^{2\pi} dy f(y)
  \cot\frac{1}{2}(y-x),
\end{equation}
with $\mathcal{P}$ denoting the Cauchy principal value integral. We
regularize currents via analytic ordering
\cite{Bazhanov:1994ft,Dymarsky:2019iny}, as described in appendix
\ref{sec:deriv-integr-moti}. The first three non-trivial charges are given by
\begin{equation}
  \begin{aligned}
    & Q_{1}=L_{0}+2\sum_{k=1}^{\infty}a_{-k}a_{k}+a_{0}^{2}-\frac{c+1}{24},\\[10pt]
 &  Q_{2} = \sum_{n\neq0}L_{-n}a_{n} + 2iQ\sum_{n>0}|n|a_{-n}a_{n} + \frac{1}{3}\sum_{\substack{n+m+p=0}} :a_{n}a_{m}a_{p}:+
  a_{0}\left(L_{0}-\frac{c+1}{24}\right),\\[10pt]
  & \begin{multlined}
      Q_{3}
  = 2\sum_{n>0}L_{-n}L_{n} + L_{0}^{2}
     -\frac{c+5}{12}L_{0} + \frac{5c^{2}+42c+37}{2880}+
     6\sum_{m\neq 0}\sum_{n+p=m} L_{-m}a_{n}a_{p}+ \\[5pt]
 +  6\left( L_{0}-\frac{c+1}{24} \right)
     \left(
     2\sum_{n>0}a_{-n}a_{n}
     + a_{0}^{2} 
     \right)
     +6iQ\sum_{n\neq 0} |n|L_{n}a_{-n}+\\[5pt]
 +  6iQ\sum_{m+n+p=0}|p|:a_{m}a_{n}a_{p}: + 2(1-5Q^{2})
     \sum_{n>0}n^{2}a_{-n}a_{n}+ \sum_{m+n+p+q=0}:a_{m}a_{n}a_{p}a_{q}:.
  \end{multlined}
\end{aligned}
\label{eq:bo2-operators}
\end{equation}
The eigenvalues of the charges \eqref{eq:bo2-operators} have the
generic form 
\begin{equation}
  \label{eq:qbo2-energies-decomposition}
  E_{I}^{(k)}(P,a) = E^{(k)}(P,a) +
  e_{I}^{(k)}(P,a)\quad (k\geq 0),
\end{equation}
where we call $E^{(k)}(P,a)$ the primary part, polynomials in
$\Delta(P)$ and $a$, while $e_{I}^{(k)}(P,a)$ is the descendant part,
being zero for primary states.  The first four eigenvalues with
$a = 0$ were presented explicitly in \cite{Litvinov:2013zda} and here
we extend that result to non-zero $a$. The descendant part can be
written in terms of the \emph{charged} Calogero-Sutherland energies
defined as
\begin{equation}
 \label{eq:calogero1}
    h_{(\lambda,\mu)}^{(k)}(P,a) = h_{\lambda}^{(k)}(P,a)+h_{\mu}^{(k)}(-P,a)
\end{equation}
with
\begin{equation}
 \label{eq:calogero2}
  h_{\lambda}^{(k)}(P,a) = b^{1-k}\sum_{j>0}\left\{\left[b(P-ia)-\frac{b^{2}}{2}
      +\lambda_{j}+jb^{2}\right]^{k}-
    \left[b(P-ia)-\frac{b^{2}}{2}+jb^{2}\right]^{k}\,\right\}.
\end{equation}
The first eigenvalue is trivial
$E^{(0)}_{I}(P,a) = E^{(0)}(a) = a$, while the first three non-trivial
eigenvalues with nonzero $a$ are
\begin{equation}
  \label{eq:qbo2-energies}
  \begin{aligned}
     &E^{(1)}_{(\lambda,\mu)}(P,a) = E^{(1)}(P,a)+h_{(\lambda,\mu)}^{(1)}(P,a),\\[5pt]
     &E^{(2)}_{(\lambda,\mu)}(P,a)=
     E^{(2)}(P,a)+\frac{i}{2}h_{(\lambda,\mu)}^{(2)}(P,a),
     \\[5pt]
     &E^{(3)}_{(\lambda,\mu)}(P,a)= E^{(3)}(P,a)
    -2h_{(\lambda,\mu)}^{(3)}(P,a) -\frac{1+b^{2}}{2}h_{(\lambda,\mu)}^{(1)}(P,a),
  \end{aligned}
\end{equation}
where
\begin{equation}
  \label{eq:primary-energies}
  \begin{aligned}
    E^{(1)}(P,a) &=\Delta(P)+a^{2} -\frac{c+1}{24},\\[5pt]
    E^{(2)}(P,a) &= a\left(\Delta(P) -\frac{c+1}{24}\right)+\frac{1}{3}a^{3},\\[5pt]
  E^{(3)}(P,a) &= \Delta(P)^2-\frac{c+5}{12}\Delta(P) +\frac{5 c^2+42c+37}{2880}+ 6a^{2}\left(\Delta(P) -\frac{c+1}{24}\right)+a^{4}.
  \end{aligned}
\end{equation}
The above eigenvalues were verified using computer algebra.

Given the known results for the $a=0$ eigenvalues in the literature
\cite{Alba:2010qc,Litvinov:2013zda,Feigin:2017gcv}, we expect that the
descendant part for $k > 3$ can be written as a linear combination of
the $h^{(q)}_{(\lambda,\mu)}(P,a)$ with $q \leq k$, although the
explicit coefficients are not known in general. This should come as a
consequence of the eigenvalues being symmetric polynomials in the
Bethe roots described in \cite{Litvinov:2013zda,Feigin:2017gcv}. An
informal argument for this structure comes from the fact that the
qBO$_{2}$ charges can be written in terms of CS Hamiltonians
\cite{estienne2012conformal}. Moreover, for the fixed $k$ eigenvalue,
the highest partition contribution $\sum_{i}\lambda_{i}^{k}$ comes
from the highest power in the current $J^{k+1}$, which sets that only
the $h^{(q)}_{(\lambda,\mu)}(P,a)$ with $q \leq k$ contribute.

\subsubsection{Spectrum Domains and Equilibrium}
\label{sec:herm-equil}

The qBO$_{2}$ energies \eqref{eq:qbo2-energies} are generically
complex, depending on the domain of $P$ and $a$. Complex eigenvalues
in principle correspond to some damping or forcing factor in the
Hamiltonian time evolution and thus can model non-equilibrium
phenomena in open quantum systems. Therefore, we do not expect
imaginary contributions in equilibrium ensembles of an isolated
quantum system. This is a physical caveat to the study of the
qBO$_{2}$ GGE ensemble for chiral CFTs like Warped CFTs. However,
complex qBO$_{2}$ eigenvalues can make sense at equilibrium for
non-chiral CFTs if the sum of holomorphic and anti-holomorphic charges
cancels their imaginary parts. That will be our working hypothesis in
the rest of this paper. In fact, in
sec.~\ref{sec:gener-gibbs-ensemble} we will compute the semiclassical
free energy and cancel its imaginary part \emph{a posteriori}, as a
result of this assumption. We will then check under what conditions
the free energy is bounded from below. In the following, we shall
analyze if and under what conditions the qBO$_{2}$ charges $Q_{2}$ and
$Q_{3}$ can be made real and bounded and discuss implications for
chiral CFTs.

To begin with, it is reasonable to assume that $P$ and $a$ are either
real or purely imaginary, as the CFT energy $E^{(1)}_{I}(P,a)$ is real
in any of those cases.  The Hermiticity condition
\eqref{eq:hermiticity-condition} is compatible with $a$ real, but we
could also consider $a$ purely imaginary by assuming
$a^{\dagger}_{n} = -a_{-n}$. This can  generate a unitary spectrum
if we restore the Kac-Moody parameter $\kappa$ and let
$\kappa\rightarrow -\kappa$, as discussed, for example, in the Warped
CFT case \cite{Apolo:2018eky}. Therefore, let us  consider the
four possibilities for the pair $(P,a)$, in which they are either real
or imaginary separately, and see what this implies for the higher
qBO$_{2}$ eigenvalues.

The simplest case to consider is that both $P$ and $a$ are real. Then,
the imaginary parts of the eigenvalues come from terms with an odd
number of Hilbert transforms in the qBO$_{2}$ charges
\eqref{eq:bo2-operators} and the imaginary coefficients in the AFLT
basis \eqref{eq:aflt-basis} (see \cite{Alba:2010qc} below their
eq.~(2.14) for explicit expressions). These result in imaginary parts
in \eqref{eq:qbo2-energies}, which cannot be cancelled in chiral CFTs.

If $P$ is imaginary and $a$ is either real or imaginary, the
conclusions are the same and we still have imaginary parts for generic
partitions. However, if $P$ is real and $a$ is purely imaginary, the
Calogero energies \eqref{eq:calogero2} are real, as the AFLT
coefficients in \eqref{eq:aflt-basis} can all be made real. On the
other hand, the primary parts of the even $k$ charges have only odd
powers of $a$ and are thus purely imaginary. As we can see from
\eqref{eq:qbo2-energies}, the $k=2$ descendant part is also purely
imaginary. This charge can thus be made real if we let
$\eta_{2}\rightarrow i\eta_{2}$ (this was effectively done in
\cite{Melnikov:2018fhb}). This pattern seems to be present for any
even $k$ and, therefore, the case $P$ real and $a$ imaginary
apparently is the only one in which all qBO$_{2}$ charges can be made
real by properly choosing the chemical potentials. This thus seems to
be the ideal case for the qBO$_{2}$ equilibrium ensemble in chiral
CFTs. Although $P$ real is incompatible with a unitary thermodynamic
limit in which $P \gg \ell$, with $\ell$ the size of the cylinder, as
this would generate negative values of $\Delta$, it could make sense
if $P$ only scales with the central charge and not with $\ell$. It
would be interesting to further investigate this possibility for
chiral CFTs in the future.

Clearly, we can still get real total charges in any domain if we sum the
holomorphic charges with their anti-holomorphic counterparts.  Let us
exemplify this by first analyzing the imaginary part of
$E^{(2)}_{(\lambda,\mu)}$ in \eqref{eq:qbo2-energies}. Expanding
\eqref{eq:calogero2} for $k=2$, we get
\begin{align}
  \label{eq:2}
   h_{\lambda}^{(2)}(P,a) &= b^{-1}\sum_{j} \lambda_{j}^{2} + 2(P-ia)|\lambda| +
                           b\sum_{j}(2j-1)\lambda_{j} \nonumber\\[5pt]
  &= b^{-1}\sum_{j}\lambda_{j}^{2} + 2(P-ia)|\lambda| + b\sum_{j}\lambda_{j}'^{2},
\end{align}
where we used the identity (see appendix \ref{sec:comb-ident})
\begin{equation}
  \label{eq:comb-ident-1}
  \sum_{j}(2j-1)\lambda_{j} = \sum_{j}\lambda_{j}'^{2},
\end{equation}
with $\lambda'$ being the conjugate partition of $\lambda$. Then, from
\eqref{eq:calogero1} and \eqref{eq:qbo2-energies}, we
have
\begin{multline}
  \label{eq:10}
  \Im E^{(2)}_{(\lambda,\mu)} =\Im E^{(2)}+\\[5pt]
  +\frac{1}{2} \left[ b^{-1}\sum_{j}(\lambda_{j}^{2}+\mu_{j}^{2}) +
    2(\Im a+\Re P)|\lambda| + 2 (\Im a - \Re P)|\mu| +
    b\sum_{j}(\lambda_{j}'^{2}+\mu_{j}'^{2}) \right].
\end{multline}
If the descendant contribution is diagonal between the sectors, the
sum of both sectors cancels the imaginary part in the GGE partition
function. More explicitly, if $\eta_{2}$ is real, the total real
eigenvalue is
\begin{equation}
  \label{eq:13}
 \eta_{2}(E^{(2)}_{(\lambda,\mu)}+\bar{E}^{(2)}_{(\lambda,\mu)})  = 2\eta_{2}
  \left[
    \Re E^{(2)}+ (\Re a - \Im P)|\lambda| + (\Re a + \Im P)|\mu|
  \right],
\end{equation}
and if $\eta_{2} \rightarrow i\eta_{2}$, the real eigenvalue is actually
\eqref{eq:10}.  Meanwhile, the charge $Q_{3}$ has a real primary part $E^{(3)}(P,a)$
in any case and the potential imaginary part comes from the $k=3$
charged Calogero energy
\begin{equation}
  \label{eq:14}
  \begin{multlined}
    h^{(3)}_{(\lambda,\mu)}(P,a) =
    b^{-2}\sum_{j}(\lambda_{j}^{3}+\mu_{j}^{3}) + 3P \left(
      b^{-1}\sum_{j}(\lambda_{j}^{2}-\mu_{j}^{2})+
      b\sum_{j} (\lambda_{j}'^{2}-\mu_{j}'^{2})  \right)-\\[5pt]
    -3ia \left( b^{-1}\sum_{j}(\lambda_{j}^{2}+\mu_{j}^{2}) +b\sum_{j}
      (\lambda_{j}'^{2}+\mu_{j}'^{2}) \right)
    + 3 
    \left(
      P^{2}-a^{2}- \frac{b^{2}}{12}
    \right)(|\lambda|+|\mu|) -\\[5pt]
    -6iPa(|\lambda|-|\mu|)+
    \frac{3}{2}\sum_{j}(2j-1)(\lambda_{j}^{2}+\mu_{j}^{2})+
    b^{2}\sum_{j}(\lambda_{j}'^{3}+\mu_{j}'^{3}), 
  \end{multlined}
\end{equation}
where we used \eqref{eq:comb-ident-1} and (see appendix \ref{sec:comb-ident})
\begin{equation}
  \label{eq:comb-ident-2}
  \sum_{j}(2j-1)^{2}\lambda_{j} = \frac{4}{3}\sum_{j}\lambda_{j}'^{3}
  - \frac{1}{3}|\lambda|.
\end{equation}
If we add the extra descendant part of $Q_{3}$ in
\eqref{eq:qbo2-energies}, there is a small cancellation of the term
$-\frac{b^{2}}{4}(|\lambda|+|\mu|)$ above and we finally get for the total
descendant part
\begin{equation}
  \label{eq:q3-descendant-part}
  \begin{multlined}
    e_{(\lambda,\mu)}^{(3)}(P,a) =
    -2b^{-2}\sum_{j}(\lambda_{j}^{3}+\mu_{j}^{3}) - 6P \left(
      b^{-1}\sum_{j}(\lambda_{j}^{2}-\mu_{j}^{2})+
      b\sum_{j} (\lambda_{j}'^{2}-\mu_{j}'^{2})  \right)+\\[5pt]
    6ia \left( b^{-1}\sum_{j}(\lambda_{j}^{2}+\mu_{j}^{2}) +b\sum_{j}
      (\lambda_{j}'^{2}+\mu_{j}'^{2}) \right)
    - 6 
    \left(
      P^{2}-a^{2}+ \frac{1}{12}
    \right)(|\lambda|+|\mu|) +\\[5pt]
    +12iPa(|\lambda|-|\mu|)-
    3\sum_{j}(2j-1)(\lambda_{j}^{2}+\mu_{j}^{2})-
    2b^{2}\sum_{j}(\lambda_{j}'^{3}+\mu_{j}'^{3}).
  \end{multlined}
\end{equation}
We can get a real eigenvalue by summing both sectors as in the
previous case.

Even if we solve the reality problem, there is still a potential
stability problem for the equilibrium state for both eigenvalues
above, as the real and imaginary parts are not necessarily bounded
from below. This could be solved by adding the other qBO$_{2}$ charges
or imposing restrictions on the spectrum, but it is still an open
problem in general. This is similar to what happens in the grand
canonical ensemble of a free gas, the partition function is only
convergent for certain values of the energy and chemical potentials,
but here we have a problem of increasingly complexity as we add more
charges. In the next section, we will make a semiclassical
saddle-point analysis and then find the free energy boundedness
conditions \emph{a posteriori}.

\section{Generalized Gibbs Ensemble of the qBO$_2$ Hierarchy}
\label{sec:gener-gibbs-ensemble}

In this section, we derive the GGE partition function of the qBO$_{2}$
hierarchy and the free energy in the thermodynamic semiclassical
limit. We also obtain a simplified form of the finite $c$ partition
function when only $\eta_{2}$ is non-zero. First, let us review the
usual CFT construction. The grand canonical partition function of a 2D
Euclidean CFT on a cylinder of size $2\pi\ell$, inverse temperature
$\beta$, angular potential $\theta$ and U(1) charge potential $\eta$
can be written as (see \cite{Kraus:2006wn}, for example)
\begin{align}
  \label{eq:cft_partition_fct}
  Z(\tau,\eta)= \Tr
  \left(
  e^{-\beta H+ i\theta \mathcal{M}}e^{ i\eta \mathcal{J}} 
  \right)
  = \Tr
  \left(
   q^{L'_{0}-\frac{c'}{24}}\bar{q}^{\bar{L}'_{0}-\frac{c'}{24}}e^{i\eta\mathcal{J}}
  \right),
\end{align}
where $q=e^{2\pi i\tau }$, with the torus moduli parameter defined as
$\tau = \frac{1}{2\pi\ell}(\theta +i\beta)$, and the total Hamiltonian
$H$, angular momentum $\mathcal{M}$ and U(1) charge $\mathcal{J}$ given by
\begin{align}
  \label{eq:85}
 H=\frac{1}{\ell}\left( L'_{0}+ \bar{L}'_{0}
  -\frac{c'}{12}\right),\quad
  \mathcal{M} =
  \frac{ 1}{\ell} 
  \left( 
    L'_{0}-\bar{L}'_{0} 
  \right),\quad \mathcal{J} = a_{0}-\bar{a}_{0},
\end{align}
and
\begin{equation}
  \label{eq:91}
  L_{0}' \equiv L_{0} + a_{0}^{2},\quad c' = c+1.
\end{equation}
This last definition is frequently used in the charged CFT literature,
which uses the Virasoro modes of the combined energy-momentum tensor
$T' = T + \frac{1}{\kappa}J^{2}$ (the Kac-Moody level is $\kappa=1$ in
our conventions).  In the following, we will discuss only one chiral
sector of the partition function and assume the same behaviour for the
other sector, which is left implicit in our equations.

The descendant basis
\eqref{eq:descendant-basis} is neither orthogonal nor normalized, but
we must define the partition function as a normalized trace to reflect
only the states' degeneracy in its sum.  To achieve this, the explicit
form of the trace in \eqref{eq:cft_partition_fct} needs to include the
inverse of the Kac matrix \eqref{eq:81} as follows
\begin{align}
  \label{eq:cft-partition-fct-explicit}
  Z(\tau,\eta)
  &= \sum_{P,a,I}\, (M^{-1})^{I,I}\,{}^{\phantom{*}}_{I}
    \langle v_{P,a}| e^{2\pi i\tau (L'_{0}-\frac{c'}{24})+
    i\eta\, a_{0} } |v_{P,a}\rangle^{\phantom{*}}_{I}
    \nonumber \\[5pt]
  &=\sum_{P,a} e^{2\pi i\tau(\Delta'-\frac{c'}{24})+ i\eta a}
    \sum_{N=|I|} e^{2\pi i \tau N}\sum_{I\in
    \mathbb{Y}_{N}}(M^{-1})^{I,I}\,M_{I,I}^{\phantom{-1}}
    \nonumber \\[5pt]
  &= \sum_{P,a,N} p_{2}'(N)\,e^{2\pi i\tau(\Delta'-\frac{c'}{24}+N)+ i\eta a},
\end{align}
where the primary eigenvalue of $L_{0}'$ is
\begin{equation}
  \label{eq:8}
  \Delta'(P,a)  \equiv \Delta(P) + a^{2}
\end{equation}
and $p'_{2}(N)$ is the number of partitions of an integer $N$ split
into two sub-partitions. The prime in $p'_{2}(N)$ means we must exclude
the trivial $L_{-1}^{N}$ descendants of the vacuum state $P=\pm Q/2$
\cite{Maloney2010,Keller2015}.

Now let us find the GGE partition function
\eqref{eq:partition-function} by including all the qBO$_{2}$
charges. As noticed in sec.~\ref{sec:bo2-hierarchy}, these charges are
diagonal in the AFLT basis. Let us explicitly show how to write the
GGE partition function in the AFLT basis by starting from the standard
descendant basis
\begin{align}
  \label{eq:gge-partition-function}
  Z(\bm{\eta}) 
  &= \sum_{P,a,I}\, (M^{-1})^{I,I}\,\,{}^{\phantom{*}}_{I}
    \langle v_{P,a}|  e^{\sum_{k}\eta_{k}Q_{k}}
    |v_{P,a}\rangle^{\phantom{*}}_{I}
    \nonumber \\[5pt]
  &=
    \sum_{P,a,I}
    \sum_{J,K}(C^{-1})^{J}_{I}(M^{-1})^{I,I}(C^{-1})^{K}_{I}\;\;
    \dualprim{J} e^{\sum_{k}\eta_{k}Q_{k}} \prim[w]{K}  
    \nonumber\\[5pt]
  &= \sum_{P,a,J}
    (\mathcal{N}_{J})^{-1}\mathcal{N}^{\phantom{1}}_{J}
    \;e^{\sum_{k}\eta_{k}E^{(k)}_{J}(P,a)
    }
    \nonumber\\[5pt]
  &=\sum_{P,a,J} 
    e^{\sum_{k}\eta_{k}E_{J}^{(k)}(P,a)},
\end{align}
where we used \eqref{eq:aflt-basis} and \eqref{eq:dual-aflt-basis} in
the second line and \eqref{eq:aflt-orthogonality} and the inverse of
\eqref{eq:diagonalization-kac-matrix} in the third line. Clearly, we
would get the same result if we calculated the trace using the
normalized AFLT basis from the beginning.

A similar argument can be used to show that the GGE partition function
is the same if we use either the commuting basis in terms of $T$ and
$J$ modes or the composite basis in terms of $T' = T +J^{2}$ and $J$
modes. It is not difficult to convince oneself that the descendants in
terms of $L'$ and $a$ operators can be written as linear combinations
of the $L$ and $a$ descendants, i.e.,
\begin{equation}
  \label{eq:15}
  \prim[v'][P',a]{I}=
  \sum_{J\in \mathbb{Y}_{|I|}}
  R_{I,J}\prim[v][P,a]{J},\quad  \dualprim[v'][P',a]{I} =
  \sum_{J\in \mathbb{Y}_{|I|}} \dualprim[v][P,a]{J}(R^{T})_{I,J}, 
\end{equation}
such that the inner product matrices in the two basis are related by
\begin{equation}
  \label{eq:39}
  M'(\Delta') = R^{T}\cdot M(\Delta)\cdot R. 
\end{equation}
Using these results and taking the inverse of \eqref{eq:39}, we can
show that
\begin{align}
  \label{eq:40}
   \sum_{P',a,I}\,(M'^{-1})_{I,I}\,\,{}^{\phantom{*}}_{I}
    \langle v'_{P',a}|  e^{\sum_{k}\eta_{k}Q_{k}}
  |v'_{P',a}\rangle^{\phantom{*}}_{I}  =
  \sum_{P,a,I}\,(M^{-1})_{I,I}\,\,{}^{\phantom{*}}_{I}
    \langle v_{P,a}|  e^{\sum_{k}\eta_{k}Q_{k}}
    |v_{P,a}\rangle^{\phantom{*}}_{I},  
\end{align}
which establishes that the normalized partition function does not
depend on such change of basis.

\subsection{Thermodynamic Limit}
\label{sec:thermodynamic-limit}

In this section, we analyze the partition functions above in
the thermodynamic limit $\ell\rightarrow\infty$ of the spectrum
defined by
\begin{equation}
  \label{eq:thermodynamic_limit}
  \Delta \sim \ell^{2},
  \quad N \sim \ell^{2},\quad a\sim \ell,
\end{equation}
where $N \equiv |I|$ refers to the descendant level.  The limit
\eqref{eq:thermodynamic_limit} is consistent with a saddle-point
computation of the CFT free energy in the large $\ell$ limit
\cite{Dymarsky:2018lhf}, using the charged Cardy formula
\cite{Cardy:1986ie,Kraus:2006wn,Pal:2020wwd} for the asymptotic
density of states $\rho(\Delta',a)$
\begin{equation}
  \label{eq:charged_cardy}
  \log\rho(\Delta',a) \sim 
  2\pi\sqrt{\frac{c'}{6}
    \left(
      \Delta' - \frac{c'}{24} -a^{2}
    \right)} =     2\pi\sqrt{\frac{c+1}{6}
    \left(
      \Delta - \frac{c+1}{24} 
    \right)}
  \sim    2\pi\sqrt{\frac{c+1}{6}
     \Delta }
\end{equation}
and the asymptotic number of 2-colored partitions
\cite{huang1970ultimate,carlip2000logarithmic}
\begin{equation}
  \label{eq:86}
  \log p_{2}(N) \sim 2\pi \sqrt{\frac{N}{3}}.
\end{equation}
The charged Cardy formula can be derived assuming that the partition
function transforms covariantly under a modular transformation
\cite{Kraus:2006wn,Dyer:2017rul}
\begin{equation}
  \label{eq:modular_covariance}
  Z
  \left(
    \frac{A\tau+B}{C\tau+D},\frac{\eta}{C\tau+D}
  \right) = e^{\frac{\pi i}{8\pi^{2}}\frac{C \eta^{2}}{C\tau+D}}Z(\tau,\eta).
\end{equation}
The Cardy formula has been rigorously derived recently in
\cite{Mukhametzhanov:2019pzy,Pal:2019zzr,Pal:2020wwd}, both in the
charged and uncharged cases. These works  remark that the Cardy
asymptotics \eqref{eq:charged_cardy} does not distinguish between
primaries and descendants. In addition, to get the asymptotics of
non-vacuum primaries, \emph{we should replace $c$ by $c-1$ in
  \eqref{eq:charged_cardy}}
\cite{Mukhametzhanov:2019pzy,Pal:2020wwd}. We shall use this
below. Moreover, \cite{Pal:2019zzr} shows that the subleading
corrections to \eqref{eq:charged_cardy} can be sensitive on how the
Cardy limit is taken for non-zero spin operators. In this paper, we
will only consider the leading order in the thermodynamic limit, so
that these subleading terms will not be relevant to our
analysis. Therefore, we will effectively assume $\theta=0$ in
\eqref{eq:cft_partition_fct}.

To derive the free energy $F_{0}$ in the thermodynamic limit
\eqref{eq:thermodynamic_limit}, we can replace the partition function
sums in \eqref{eq:cft-partition-fct-explicit} by integrals
\begin{align}
  \label{eq:3}
  Z(\tau,\eta) \sim \int  da\,d\Delta\, dN\, e^{2\pi
  \sqrt{\frac{c}{6}\Delta
  }}
  e^{2\pi
  \sqrt{\frac{N}{3}}}e^{-\frac{\beta}{\ell}(\Delta+N+a^{2})+i\eta a}= e^{F_{0}},
\end{align}
where we replaced the sum over $P$ by an integral over $\Delta$, which
is simpler in this case.  In the saddle-point approximation for large
$\Delta$, $a$ and $N$, we get the saddle-point values
\begin{equation}
  \label{eq:87}
  \Delta_{*} = \frac{c\pi^{2}\ell^{2}}{6\beta^{2}},\quad a_{*} =
  \frac{i\eta \ell}{2\beta},\quad N_{*} = \frac{\pi^{2}\ell^{2}}{3\beta^{2}}
\end{equation}
and the free energy is then
\begin{equation}
  \label{eq:92}
  F_{0} = \frac{(c+2)\pi^{2}\ell}{6\beta} -\frac{\eta^{2}\ell}{4\beta}.
\end{equation}
We recover the usual Virasoro CFT result by making $\eta=0$ and
$c\rightarrow c-2$, whose origin comes from the shifted Virasoro
vacuum energy with $c'= c+1$ and the density of descendants which also
includes the U(1) descendants. Notice that, if $\eta$ is real, $a_{*}$ is
imaginary in \eqref{eq:87}. This is not a problem \emph{per se}, but
$F_{0}$ is then not bounded from below. Therefore, we will consider
that $\eta$ is purely imaginary for the rest of this work.

Let us now consider the thermodynamic limit of the qBO$_{2}$ charges.
As we expressed the GGE partition function
\eqref{eq:gge-partition-function} as a sum over $P$, we should
consider the thermodynamic scaling of $P$. In the Liouville branch,
$P=ip$ and we will assume that
\begin{equation}
  \label{eq:liouville-thermo-limit}
  p \sim 
 \frac{\ell}{b^{2}},\quad a\sim \frac{\ell}{b^{2}},
\end{equation}
such that $\Delta \gg \frac{c-1}{24}$ in the thermodynamic limit. This
is consistent with \eqref{eq:thermodynamic_limit} and we also
introduced the Liouville parameter dependence to be considered in the
next section. An alternative option would be
$p \sim \ell\sqrt{(c-1)/6}\sim \ell Q \approx \ell/b^{2}$, but, as the
qBO$_{2}$ descendant eigenvalues \eqref{eq:calogero2} are explicitly
written in terms of $b$, it is much more convenient to consider the
semiclassical expansion $b\rightarrow 0$ using
\eqref{eq:liouville-thermo-limit}.

The scaling \eqref{eq:liouville-thermo-limit} implies that the primary
parts scale as $E^{(k)}(P,a) \sim \ell^{k+1}$ in the thermodynamic
limit and lower order terms in \eqref{eq:primary-energies} can be
discarded. With respect to the descendant parts
$e_{(\lambda,\mu)}^{(k)}$, following a similar argument as in
\cite{Dymarsky:2018lhf}, we consider the following effective scaling
for the partitions
\begin{equation}
  \label{eq:41}
  \sum_{j}\lambda_{j},\, \sum_{j}\lambda'_{j} \sim \ell^{2},
\end{equation}
such that, for example,
\begin{equation}
  \label{eq:17}
  \sum_{j} \lambda_{j}^{k} \sim |\lambda|^{1/2} \times |\lambda|^{k/2}
  = |\lambda|^{(k+1)/2} \sim \ell^{k+1},
\end{equation}
and
\begin{equation}
  \label{eq:129}
  \sum_{j}(2j-1)^{k}\lambda_{j} \sim \sum_{j} {\lambda'_{j}}^{k+1} \sim \ell^{k+2}.
\end{equation}
This means that all terms in the Calogero energies
\eqref{eq:calogero2} scale as
$h_{(\lambda,\mu)}^{(k)} \sim \ell^{k+1}$, just like the primary
parts, and the descendant part $e_{(\lambda,\mu)}^{(k)}$ is thus
dominated by $h_{(\lambda,\mu)}^{(k)}$. The qBO$_{2}$ eigenvalues have
thus the following structure in the thermodynamic limit
\begin{equation}
  \label{eq:16}
  E^{(k)}_{I}
  \sim \sum_{q=0}^{\lfloor\frac{k+1}{2}\rfloor}f_{q}^{(k)}a^{k+1-2q}\Delta^{q}+
  d^{(k)}h_{I}^{(k)}(P,a),
\end{equation}
with $f^{(k)}_{q} ,d^{(k)}$ being constants.  Ideally, we would like
to integrate the descendant parts in the partition function and then
do a saddle-point calculation of the resulting primary
character. Finding the coefficients $f_{q}^{(q)}$ and $d^{(k)}$
requires knowing the coefficients of the qBO$_{2}$ eigenvalues for any
$k$, which we do not currently know. Apart from this, the complete
integration of the partition function might be done in certain special
cases, like when only a finite number of chemical potentials is
non-zero. Below we are going to discuss the case when only $\eta_{2}$
and $\eta_{3}$ are non-zero.

\subsection{GGE Free Energy in the Thermodynamic Semiclassical Limit}
\label{sec:saddle-point-semicl}

Here we calculate the free energy of the qBO$_{2}$ GGE in the
semiclassical limit. For a generic non-relativistic critical system,
the leading saddle-point could be different from the CFT one, but we
will not consider this possibility. We shall proceed by considering
the extra qBO$_{2}$ charges as corrections to the CFT saddle-point
free energy, as in
\cite{Dymarsky:2018lhf,Maloney:2018yrz,Dymarsky:2018iwx,Dymarsky:2019etq},
equivalent to assuming that the chemical potentials are small relative
to the temperature.

As we saw above, the eigenvalues $E^{(k)}_{I}$ scale as $\ell^{k+1}$
in the thermodynamic limit. Therefore, we shall rescale the chemical
potentials as $\eta_{k} \rightarrow \frac{1}{\ell^{k}}\eta_{k}$, such
that all the charges contribute at the same order in $\ell$ in the GGE
partition function.  With respect to the semiclassical limit of the
qBO$_{2}$ eigenvalues, the primary part scales as $b^{-k-1}$ and the
descendant part scales as $b^{-k+1}$ (or order one and order $b^{2}$
after an overall rescaling by $b^{k+1}$). Therefore, at leading order
in $c$, the GGE partition function depends only on the primary
parts. In particular, if we set $a=0$ the saddle-point calculation
gives the qKdV result
\cite{Dymarsky:2018lhf,Maloney:2018yrz,Dymarsky:2018iwx,Dymarsky:2019etq},
as $E^{(2n)} = 0$ and $E^{(2n+1)} \sim \Delta^{n+1}$.  In the
following, we present the saddle-point computation including the U(1)
charge with the leading $b^{2}\approx 1/c$ correction.

In the thermodynamic semiclassical limit, the GGE partition function
\eqref{eq:gge-partition-function} can be approximated by 
\begin{equation}
  \label{eq:20}
  Z(\bm{\eta}) = \int d\Delta da\,
  e^{\mathcal{L}(\Delta,a,\bm{\eta})} = e^{F(\bm{\eta})}, 
\end{equation}
with
\begin{equation}
  \label{eq:23}
  \mathcal{L} = 2\pi\sqrt{\frac{c}{6}\Delta}+ \eta a -
  \frac{\beta}{\ell}(\Delta+a^{2})
  +\frac{\eta_{2}}{\ell^{2}}(a\Delta + \frac{1}{3}a^{3})
  +\frac{\eta_{3}}{\ell^{3}}(\Delta^{2}+6a^{2}\Delta+a^{4})+\dots.
\end{equation}
Notice that we rotated $\eta\rightarrow i\eta$ and rescaled the
chemical potentials with respect to $\ell$, as discussed in the
previous subsection. Furthermore, similarly to
\cite{Dymarsky:2018lhf}, we rescale the parameters in \eqref{eq:23} as
\begin{equation}
  \label{eq:31}
  a = 
  \frac{\pi\ell}{b\beta}\alpha,\quad
  P = 
  \frac{i\pi\ell}{b\beta}p \quad\Rightarrow\quad
  \Delta \sim
  \left(
    \frac{\pi\ell}{b\beta}p
  \right)^{2},
\end{equation}
and thus
\begin{equation}
  \label{eq:32}
  \mathcal{L}^{(0)} \equiv \frac{\mathcal{L}}{f_{0}} =
  2p +\tilde{\eta}\alpha-(p^{2}+\alpha^{2}) +\tilde{\eta}_{2}(\alpha
  p^{2}+\frac{1}{3}\alpha^{3})+\tilde{\eta}_{3}(p^{4}+6\alpha^{2}p^{2}+\alpha^{4})
  + \dots,
\end{equation}
where we defined
\begin{equation}
  \label{eq:82}
  f_{0} =\frac{\pi^{2}\ell}{\beta b^{2}},\quad
  \tilde{\eta} = 
    \frac{b}{\pi}
  \eta,\quad
  \tilde{\eta}_{k} = 
 \left(
    \frac{\pi}{b}
  \right)^{k-1}\frac{\eta_{k}}{\beta^{k}},\, k>1.
\end{equation}
In the following, we shall consider the free energy as a function of
the $\tilde{\eta}_{k}$ instead of $\eta_{k}$, consistent with our
assumptions on the thermodynamic semiclassical limit. Notice that the
overall constant $f_{0}$ in \eqref{eq:32} is just \eqref{eq:92} in the
semiclassical limit $b\rightarrow 0$ and we used
\eqref{eq:liouville-notation} to approximate
\begin{equation}
  \label{eq:42}
 \sqrt{ \frac{c}{6}}=\frac{1}{b}\sqrt{1+\frac{13}{6}b^{2}+b^{4}}
 \approx \frac{1}{b}.
\end{equation}
We want to find the extrema
of \eqref{eq:32} from
\begin{gather}
  \label{eq:88}
  \del[p]\mathcal{L}\big\rvert_{p_{*},\alpha_{*}} = 0,\quad
  \del[\alpha]\mathcal{L}\big\rvert_{p_{*},\alpha_{*}} = 0,
\end{gather}
and, if possible, those points that are minima of the free energy,
which requires the determinant of the Hessian matrix to be
positive. For generic values of the chemical potentials and domain of
the variables, there is no global minima. However, it is still
possible to find minima if we stick to a perturbative solution close
to the CFT saddle-point. In the following, let us consider two cases
with only $\eta_{2}$ and $\eta_{3}$ being non-zero, discarding all the
other chemical potentials.

First, suppose that $\tilde{\eta}_{k}=0$ for $k>2$ in
\eqref{eq:32}. For $\alpha$ being a real number, $\mathcal{L}$ is
clearly not bounded, as the highest order monomial is
$\alpha^{3}$, but we shall press on and find the extrema. Solving for
$p$ in \eqref{eq:88}, we get
\begin{equation}
  \label{eq:93}
  p_{*} = \frac{1}{1-\tilde{\eta}_{2}\alpha_{*}}.
\end{equation}
Plugging into the equation for $\alpha_{*}$ gives a quartic polynomial
\begin{equation}
  \label{eq:94}
  \tilde{\eta}_{2}^{3}\alpha_{*}^{4}-4\tilde{\eta}_{2}^{2}\alpha_{*}^{3}+
  \tilde{\eta}_{2}(5+\tilde{\eta}\tilde{\eta}_{2})\alpha_{*}^{2}
  -2(1+\tilde{\eta}\tilde{\eta}_{2})\alpha_{*}+\tilde{\eta}+\tilde{\eta}_{2} = 0.
\end{equation}
To analyze how many real roots we have, in general we need to consider
the discriminant of this polynomial. But if we expand the roots for
small $\tilde{\eta}_{2}$, there are two complex roots, which we
discard, and two real roots, one of which blows up as
$\tilde{\eta}_{2}$ goes to zero. Choosing the only regular real root
in this limit, we get the free energy
\begin{equation}
  \label{eq:95}
  F = f_{0}
  \left[1+\frac{\tilde{\eta}^{2}}{4}+
    \frac{1}{24}(12 \tilde{\eta}+\tilde{\eta}^{3})
    \tilde{\eta}_{2}
    +
    \frac{1}{64}(16+24\tilde{\eta}^{2}+\tilde{\eta}^{4})\tilde{\eta}_{2}^{2}+
    \mathcal{O}(\tilde{\eta}_{2}^{3})\right].
\end{equation}
Notice that if $\tilde{\eta}_{2}=0$ above, we recover the result \eqref{eq:92}.

Now, if we assume in addition that $ \tilde{\eta}_{3} > 0$, there
should be at least one global minimum for $\mathcal{L}$ as the highest
powers of $\alpha$ and $p$ in $\mathcal{L}$ are even.  There are four
roots for $p$ and $\alpha$, but only one pair is real and regular for
small $\tilde{\eta}_{2}$ and $\tilde{\eta}_{3}$. Their expansion up to
2nd order is
\begin{equation}
  \label{eq:104}
  \begin{aligned}
    p_{*} &= 1 +\frac{\tilde{\eta}}{2}\tilde{\eta}_{2}+ \left(
      2+\frac{3}{2}\tilde{\eta}^{2} \right)\tilde{\eta}_{3}+ \left(
      1+\frac{3\tilde{\eta}^{2}}{4}
    \right)\frac{\tilde{\eta}_{2}^{2}}{2} +\left( 10\tilde{\eta} +
      \frac{5}{2}\tilde{\eta}^{3}
    \right)\tilde{\eta}_{2}\tilde{\eta}_{3}+\\[5pt]
    &\quad+\left( 12+30\tilde{\eta}^{2}+\frac{15}{4}\tilde{\eta}^{4}
    \right)\tilde{\eta}_{3}^{2}+\dots
    ,\\[5pt]
    \alpha_{*} &= \frac{\tilde{\eta}}{2}
    +\frac12\left(1+\frac{1}{4}\tilde{\eta}^{2}\right)\tilde{\eta}_{2}
    +\left(3\tilde{\eta}+\frac{1}{4}\tilde{\eta}^{3}\right)\tilde{\eta}_{3}
    +\frac{1}{4}\left( 3\tilde{\eta}+\frac{1}{4}\tilde{\eta}^{3}
    \right)\tilde{\eta}_{2}^{2}+
    \\[5pt]
    &\quad+ \left(
      5+\frac{15}{2}\tilde{\eta}^{2}+\frac{5}{16}\tilde{\eta}^{4}
    \right)\tilde{\eta}_{2}\tilde{\eta}_{3} +\left(
      30\tilde{\eta}+15\tilde{\eta}^{3}+\frac{3}{8}\tilde{\eta}^{5}
    \right)\tilde{\eta}_{3}^{2}+\dots,
  \end{aligned}
\end{equation}
and the free energy expansion is then given by
\begin{equation}
  \label{eq:97}
  \begin{multlined}
    \frac{F}{f_{0}}= 1 + \frac{\tilde{\eta}^{2}}{4}
    + \left(
      \frac{\tilde{\eta}}{2}+\frac{\tilde{\eta}^{3}}{24}
    \right)\tilde{\eta}_{2}
    + \left(
      1+\frac{3 \tilde{\eta}^{2}}{2}+\frac{\tilde{\eta}^{4}}{16}
    \right)\frac{\tilde{\eta}_{2}^{2}}{4}
    +\\[5pt]
    + \left[
      1+\frac{3\tilde{\eta}^{2}}{2}+\frac{\tilde{\eta}^{4}}{16}+
      \left(
        5\tilde{\eta}+\frac{5\tilde{\eta}^{3}}{2}+
        \frac{\tilde{\eta}^{5}}{16}
      \right)\tilde{\eta}_{2}
      + \left(
        7+\frac{105}{4}(\tilde{\eta}^{2}+\frac{\tilde{\eta}^{4}}{8})
        +\frac{7\tilde{\eta}^{6}}{64}
      \right)\frac{\tilde{\eta}_{2}^{2}}{2}
    \right]\tilde{\eta}_{3}
    +\\[5pt]
    + \left[ 4+15\tilde{\eta}^{2}+\frac{15\tilde{\eta}^{4}}{4}
      +\frac{\tilde{\eta}^{6}}{16}
      + \left(
        56\tilde{\eta}+70\tilde{\eta}^{3}+\frac{21\tilde{\eta}^{5}}{2}
        +\frac{\tilde{\eta}^{7}}{8}
      \right)\tilde{\eta}_{2}+\right.\\[5pt]
    \left.+ \left( 45+315\tilde{\eta}^{2}+
        \frac{1575\tilde{\eta}^{4}}{8}+\frac{315\tilde{\eta}^{6}}{16}+
        \frac{45\tilde{\eta}^{8}}{256}
      \right)\tilde{\eta}_{2}^{2} \right]\tilde{\eta}_{3}^{2}+\dots.
  \end{multlined}
\end{equation}
If we include all the chemical potentials in the partition function,
we can in principle find a general solution in perturbative form
\begin{equation}
  \label{eq:102}
  \begin{aligned}
    p_{*} &= 1+\sum_{N=2}^{\infty}\, \sum_{m_{2},m_{3},\dots,m_{N} =
      2}p_{m_{2},
      m_{3},\dots,m_{N}}(\tilde{\eta})\,\tilde{\eta}_{m_{2}}\tilde{\eta}_{m_{3}}\dots
    \tilde{\eta}_{m_{N}} ,
    \\[5pt]
    \alpha_{*} &= \frac{\tilde{\eta}}{2}+
    \sum_{N=2}^{\infty}\,\sum_{m_{2},m_{3},\dots,m_{N} = 2}
    \alpha_{m_{2},
      m_{3},\dots,m_{N}}(\tilde{\eta})\,\tilde{\eta}_{m_{2}}
    \tilde{\eta}_{m_{3}}\dots \tilde{\eta}_{m_{N}},
    \\[5pt]
    \frac{F}{f_{0}} &=1+\frac{\tilde{\eta}^{2}}{4}
    +\sum_{N=2}^{\infty}\,\sum_{m_{2},m_{3},\dots,m_{N} = 2}
    f_{m_{2}, m_{3},\dots,m_{N}}(\tilde{\eta})\,
    \tilde{\eta}_{m_{2}}\tilde{\eta}_{m_{3}}\dots
    \tilde{\eta}_{m_{N}}.
  \end{aligned}
\end{equation}
We could not find the generic expansion coefficients here as in
\cite{Maloney:2018yrz}, as we would need to know the primary energies
for any $k$ and those have a more complicated form than the qKdV ones.
For the particular case where only $\tilde{\eta}_{2}$ and
$\tilde{\eta}_{3}$ are non-zero, if we substitute such series
solutions into \eqref{eq:88}, we find that all coefficients are
positive if all the chemical potentials are positive in our
conventions. This in particular implies that the free energy is
positive and bounded from below. The same computation is valid for the
anti-holomorphic sector, as the eigenvalues are all real in this case.

\subsection{Leading Quantum Correction}
\label{sec:1c-correction}

To calculate the free energy with the leading quantum correction, we
need to consider the sum over descendants in the partition function
\begin{equation}
  \label{eq:105}
  Z(\bm{\eta}) = \sum_{I}\int dp d\alpha\,
  e^{\mathcal{L}_{I}(p,\alpha,\bm{\eta})}\times (\text{\emph{anti-holo}})= e^{F(\bm{\eta})},
\end{equation}
where \emph{anti-holo} represents the anti-holomorphic sector. The
perturbative expansion of $\mathcal{L}_{I}$ is of the form
\begin{equation}
  \label{eq:106}
  \frac{\mathcal{L}_{I}}{f_{0}} = 
  \mathcal{L}^{(0)}+
  b^{2}
  \left(\mathcal{L}_{I}^{(1)}+\frac{13}{6}p\right)+\mathcal{O}(b^{4}),
\end{equation}
where $\mathcal{L}^{(0)} $ is defined in \eqref{eq:32} and only the
subleading correction $\mathcal{L}^{(1)}_{I}$ depends on the
descendant part. Notice that we also considered the contribution from
\eqref{eq:42} in the subleading correction. This is equivalent to the
leading correction in the central charge expansion as
$\frac{6}{c} \approx b^{2} $ for $b$ small.

As above, we will consider only $\eta_{2}$ and $\eta_{3}$ as being
non-zero. The first qBO$_{2}$ eigenvalue is given by
\begin{equation}
  \label{eq:107}
  -\frac{\beta}{\ell f_{0}}E^{(1)}_{I} \sim
  -(p^{2}+\alpha^{2})- b^{2}\tilde{N},
\end{equation}
where we assumed the scaling
\begin{equation}
  \label{eq:43}
  N \sim 
  \left(
    \frac{\pi\ell}{\beta}
  \right)^{2}\tilde{N},
\end{equation}
whose dependence on $\pi$ and $\beta$ should also be considered in the
scalings \eqref{eq:41}.  The semiclassical limit of the Calogero
energies \eqref{eq:calogero2} is easily obtained  using
\eqref{eq:41} and \eqref{eq:31}
\begin{align}
  \label{eq:108}
  h_{\lambda}^{(k)}(P,a)
  &\approx
    b^{1-k}  
    \sum_{j>0}
    \left[
    (b(P-ia)+\lambda_{j})^{k}-(b(P-ia))^{k}
    \right]\nonumber\\[5pt]
  &=  b^{1-k}  
    \sum_{j>0}\sum_{n=0}^{k-1}
    \begin{pmatrix}
      k \\ n
    \end{pmatrix} \lambda_{j}^{k-n}(b(P-ia))^{n}\nonumber\\[5pt]
  &\sim b^{2} \left( \frac{\pi\ell}{b\beta} \right)^{k+1}\sum_{m>0}
    \left[ (i(p-\alpha)-m)^{k}-(i(p-\alpha))^{k} \right]\tilde{N}_{m},
\end{align}
where we used the binomial expansion in the second line and the
thermodynamic scaling in the last line. Considering the full Calogero
contribution \eqref{eq:calogero1}, we get
\begin{equation}
  \label{eq:110}
  \begin{multlined}
    \frac{ \eta_{k}}{f_{0}\ell^{k}}e^{(k)}_{I} \sim
    b^{2}\tilde{\eta}_{k}\sum_{m} \left\{
      [(m+i(p-\alpha))^{k}-(i(p-\alpha))^{k}]\tilde{N}_{1,m}+\right.\\[5pt]
    \left.  [(m-i(p+\alpha))^{k}-(-i(p+\alpha))^{k}]\tilde{N}_{2,m}
    \right\},
  \end{multlined}
\end{equation}
which we conjecture should be valid  for any
$k$.  Thus, including all the terms \eqref{eq:110}, we get
\begin{align}
  \label{eq:111}
  \mathcal{L}_{I}^{(1)} = \sum_{j=1,2}\sum_{m>0}
  \sum_{k>0}
  d_{k}\tilde{\eta}_{k}[(m+i(\sigma_{j}p-\alpha))^{k}-
  (i(\sigma_{j}p-\alpha))^{k}]\tilde{N}_{j,m}, 
\end{align}
with $(\sigma_{1},\sigma_{2}) =(1,-1) $ and $d_{k}$ being
constants. The constants we know explicitly are
$(d_{1},d_{2},d_{3})=(1,i/2,-2)$.

We want to calculate the free energy again using the saddle-point
approximation, but now the descendant part also depends on $p$ and
$\alpha$.  We can use the leading semiclassical saddle-point
$(p_{*},\alpha_{*})$ to calculate the free energy up to $b^{2}$ order
due to the following
\begin{equation}
  \label{eq:112}
  \begin{aligned}
    \frac{1}{f_{0}}\mathcal{L}_{I} \left(p_{*}+b^{2}
      p_{1},\alpha_{*}+b^{2}\alpha_{1}, \bm{\eta}\right) &=
    \mathcal{L}^{(0)}(p_{*},\alpha_{*}, \bm{\tilde{\eta}}) +\\[5pt]
    &+ b^{2} \left(
      \mathcal{L}^{(1)}_{I}(p_{*},\alpha_{*}, \bm{\tilde{\eta}}) +
      \frac{13}{6}p_{*} \right) +\mathcal{O}(b^{4}),
  \end{aligned}
\end{equation}
where we used the saddle-point condition \eqref{eq:88} to cancel the
first correction to $\mathcal{L}^{(0)}$.  Thus,
\begin{equation}
  \label{eq:114}
  Z \sim e^{f_{0}(\mathcal{L}^{(0)}(p_{*},a_{*}, \bm{\tilde{\eta}})+\frac{13 b^{2}}{6}p_{*})}
  \sum_{I}e^{ f_{0}b^{2}\mathcal{L}^{(1)}_{I}(p_{*},\alpha_{*}, \bm{\tilde{\eta}})}\times (\text{\emph{anti-holo}}).
\end{equation}
In general, the quantum correction has an imaginary part, but, as
discussed in sec.~\ref{sec:herm-equil}, this can be canceled by
considering the anti-holomorphic contribution to the partition
function. Therefore, as we are considering a real saddle-point
$(p_{*},\alpha_{*})$ and that all chemical potentials are real, we can
simplify the quantum correction to
\begin{equation}
  \label{eq:45}
  \sum_{I}e^{
    f_{0}b^{2}\mathcal{L}^{(1)}_{I}(p_{*},\alpha_{*},
    \bm{\tilde{\eta}})}\times (\text{\emph{anti-holo}})=
  \sum_{\{\tilde{N}_{j,m}\}}
  e^{-
    \sum_{j,m}g_{j,m}(p_{*},\alpha_{*}, \bm{\tilde{\eta}})\tilde{N}_{j,m}},
\end{equation}
where $g_{j,m}$ is a real function obtained by cancelling the
imaginary part of \eqref{eq:111}.  In the particular case in which
only the two first chemical potentials are non-zero, we have
\begin{equation}
  \label{eq:121}
  \begin{multlined}
    g_{j,m} = \frac{\pi^{2}\ell}{\beta} \left\{
      2\tilde{\eta}_{3}m^{3} + \left[
        1+(\sigma_{j}p_{*}-\alpha_{*})\tilde{\eta}_{2}
        -6(\sigma_{j}p_{*}-\alpha_{*})^{2}\tilde{\eta}_{3}
      \right]m
     \right\}.
  \end{multlined}
\end{equation}
The sum over descendants in \eqref{eq:45} can be done exactly via
geometric series if we can guarantee that the sum is convergent, which
requires the exponents to be positive for all $m$. For fixed $m$, the
exponent is
\begin{align}
  \label{eq:46}
  g_{1,m}\tilde{N}_{1,m}+  g_{2,m}\tilde{N}_{2,m}
  &= \frac{\pi^{2}\ell}{\beta}
    \left\{
    \left[
    2\tilde{\eta}_{3}m^{3}+
    \left(
    1-\alpha_{*}\tilde{\eta}_{2}-6(p_{*}^{2}+\alpha_{*}^{2})\tilde{\eta}_{3}
    \right)m
    \right](\tilde{N}_{1,m}+\tilde{N}_{2,m})+ \right.\nonumber\\[5pt]
  &\left.+ p_{*}(\tilde{\eta}_{2}+12\alpha_{*}\tilde{\eta}_{3})(\tilde{N}_{1,m}-\tilde{N}_{2,m})
    \right\}\nonumber\\[5pt]
  &=\frac{\pi^{2}\ell}{\beta}
    \left\{
    2\tilde{\eta}_{3}m^{3}+
    \left[
    1-6(p_{*}^{2}-\alpha_{*}^{2})\tilde{\eta}_{3}
    \right]m
    \right\}(\tilde{N}_{1,m}+\tilde{N}_{2,m}),
\end{align}
where we assumed in the second line that
\begin{equation}
  \label{eq:51}
  \tilde{\eta}_{2}+12\alpha_{*}\tilde{\eta}_{3}=0
\end{equation}
to cancel the unbounded $\tilde{N}_{1,m}-\tilde{N}_{2,m}$
term. Furthermore, we assume that 
\begin{equation}
  \label{eq:50}
  2\tilde{\eta}_{3}[3(p_{*}^{2}-\alpha_{*}^{2})-1] \leq 1
\end{equation}
for the remaining term in \eqref{eq:46} to be positive under the
condition $\tilde{\eta}_{3}>0$. The first condition \eqref{eq:51} is
incompatible with all the chemical potentials being positive, so we
make $\tilde{\eta}_{2} \rightarrow -\tilde{\eta}_{2}$. However, in
this case we cannot guarantee that $\alpha_{*}$ and the classical free
energy $F$ are positive, as discussed before. We can check the
conditions \eqref{eq:51} and \eqref{eq:50} at leading order in the
chemical potentials
\begin{align}
  \label{eq:52}
  \tilde{\eta}_{2}=-6\tilde{\eta}\tilde{\eta}_{3},\quad
  \tilde{\eta}_{3}[4-\frac{3}{2}\tilde{\eta}^{2}] \leq 1.
\end{align}
If the free energy is positive at this order, using the first equation
in \eqref{eq:52} we have
\begin{equation}
  \label{eq:53}
  \frac{F}{f_{0}} \approx 1+\frac{\tilde{\eta}^{2}}{4}+
  \left(1-\frac{3\tilde{\eta}^{2}}{2}-\frac{9\tilde{\eta}^{4}}{16}\right)
  \tilde{\eta}_{3}\geq 0.
\end{equation}
For $\tilde{\eta}$ small enough such that
\begin{equation}
  \label{eq:55}
  4-\frac{3}{2}\tilde{\eta}^{2}> 0,
  \quad 1-\frac{3\tilde{\eta}^{2}}{2}-\frac{9\tilde{\eta}^{4}}{16}>0, 
\end{equation}
we have at least one range compatible with the conditions above
\begin{equation}
  \label{eq:54}
  -\frac{1+\frac{\tilde{\eta}^{2}}{4}}{1-\frac{3\tilde{\eta}^{2}}{2}-\frac{9\tilde{\eta}^{4}}{16}}
  <0 \leq \tilde{\eta}_{3} \leq \frac{1}{4-\frac{3}{2}\tilde{\eta}^{2}}.
\end{equation}
Considering higher-order corrections in the chemical potentials
complicates this analysis and requires a better knowledge of the
saddle-point solutions. Assuming that convergence can be guaranteed by
assumptions similar to the previous ones, we have
\begin{align}
  \label{eq:115}
  \sum_{\{\tilde{N}_{j,m}\}}
  e^{-
  \sum_{j,m}g_{j,m}(p_{*},\alpha_{*},
  \bm{\tilde{\eta}})\tilde{N}_{j,m}} =
  \prod_{j=1,2}\prod_{m=1}^{\infty}\frac{1}{1-e^{-
  g_{j,m}(p_{*},\alpha_{*}, \bm{\tilde{\eta}})}},
\end{align}
and the free energy is thus given by
\begin{equation}
  \label{eq:118}
  \begin{multlined}
    F(\bm{\tilde{\eta}})= 2f_{0}
    \tilde{\mathcal{L}}^{(0)}(\bm{\tilde{\eta}})
    +\frac{13\pi^{2}\ell}{3\beta}p_{*}(\bm{\tilde{\eta}}) -
    \sum_{j=1,2}\sum_{m>0}\log \left( 1-e^{-
        \tilde{g}_{j,m}(\bm{\tilde{\eta}})} \right) ,
  \end{multlined}
\end{equation}
where
$\tilde{\mathcal{L}}^{(0)}(\bm{\tilde{\eta}})\equiv
\mathcal{L}^{(0)}(p_{*}(\bm{\tilde{\eta}}),\alpha_{*}(\bm{\tilde{\eta}}),
\bm{\tilde{\eta}})$ and
$\tilde{g}_{j,m}(\bm{\tilde{\eta}})\equiv
g_{j,m}(p_{*}(\bm{\tilde{\eta}}),\alpha_{*}(\bm{\tilde{\eta}}),
\bm{\tilde{\eta}})$. This gives the free energy at the leading order
in $1/c$ considering both holomorphic and anti-holomorphic sectors.

There is another way to obtain a convergent descendant sum if we allow
for a purely imaginary $a$. If we make the transformations
$\tilde{\eta}\rightarrow i\tilde{\eta}$,
$\tilde{\eta}_{2}\rightarrow i\tilde{\eta}_{2}$ and
$\alpha_{*} \rightarrow i\alpha_{*}$, the resulting saddle-point
values of $p_{*}$, $\alpha_{*}$ and $F$ are still real, although in
this case not necessarily positive. The interesting point of this
choice is that the terms with odd powers of $P$ in the descendant
parts are cancelled, which mitigate part of the unboundedness problem
we had in \eqref{eq:46}.  In this case, the function $g_{j,m}$ results in
\begin{equation}
  \label{eq:56}
  g_{j,m} = \frac{\pi^{2}\ell}{\beta}
  \left\{
    2\tilde{\eta}_{3}m^{3}+
    \frac{1}{2}(\tilde{\eta}_{2}+12\alpha_{*}\tilde{\eta}_{3}) m^{2}
    +
    \left[
      1+ \alpha_{*}\tilde{\eta}_{2}-6(p_{*}^{2}-\alpha_{*}^{2})\tilde{\eta}_{3}
    \right]m
  \right\}.
\end{equation}
We could then analyze the positivity conditions for the descendant
part and the free energy as above. However, there is not much
conceptual gain of considering this case, so we will skip it here. Our
conclusion is that the conditions to guarantee convergence of the
partition function sum and the boundedness of the free energy are
delicate when we include quantum corrections. The complete problem
including all chemical potentials and quantum corrections shall be
studied elsewhere.

\subsection{Free Energy at Finite $c$ with non-zero $\eta_{2}$}
\label{sec:real-branch}

We show now how to obtain the free energy from \eqref{eq:20} at finite
$c$ when only $\eta_{2}$ is non-zero, apart from $\beta$ and $\eta$
also non-zero. First, if we consider $\eta_{2}$ to be real, the real
part of $E^{(2)}_{I}$ is given by \eqref{eq:13}. Therefore, summing
the holomorphic and anti-holomorphic charges, we have in the
thermodynamic limit \begin{multline}
  \label{eq:57}
  \frac{1}{2f_{0}}\mathcal{L} = 
  2p+\tilde{\eta}\alpha-(p^{2}+\alpha^{2})+
  \tilde{\eta}_{2}(\alpha p^{2}+\frac{1}{3}\alpha^{3})
  -b^{2}\sum_{j,m}
  \left[
  1-\tilde{\eta}_{2}(\alpha-\sigma_{j} p)
  \right]m \tilde{N}_{m}.
\end{multline}
Defining
\begin{equation}
  \label{eq:58}
  \tau_{\pm}(p,\alpha,\tilde{\eta}_{2}) \equiv \tau_{0}\left[
    1-\tilde{\eta}_{2}(\alpha\pm p)
  \right],\quad \tau_{0} = \frac{i\pi\ell}{\beta},
\end{equation}
it is clear that, if $\Im \tau_{\pm} >0$, we can easily sum up the
descendant part as geometric series
\begin{equation}
  \label{eq:109}
\prod_{m=1}^{\infty}
\left(
  \sum_{\tilde{N}_{m}=0}^{\infty}e^{2\pi i \tau_{+}m\tilde{N}_{m}}
\right)
\prod_{n=1}^{\infty}
\left(
  \sum_{\tilde{N}_{n}=0}^{\infty}e^{2\pi i \tau_{-}m\tilde{N}_{n}}
\right)
  = \prod_{n=1}^{\infty}\frac{1}{1-e^{2\pi i
      \tau_{+}n}}\prod_{m=1}^{\infty}\frac{1}{1-e^{2\pi i \tau_{-}m}}
\end{equation}
and thus
\begin{equation}
  \label{eq:124}
  \mathcal{L} =  2f_{0}
  \left[
    2p+\tilde{\eta}\alpha-(p^{2}+\alpha^{2})+
    \tilde{\eta}_{2}(\alpha p^{2}+\frac{1}{3}\alpha^{3})
  \right] -\sum_{n=1}^{\infty}\log(1-e^{2\pi i \tau_{+}n})
  -\sum_{m=1}^{\infty}\log(1-e^{2\pi i \tau_{-}m})
\end{equation}
The next step to get the free energy is to do a saddle-point
calculation, which is exact in the thermodynamic limit
\cite{Dymarsky:2018lhf}. The resulting saddle-point equations
\eqref{eq:88} are
\begin{equation}
  \label{eq:113}
  \begin{aligned}
     \frac{1}{2f_{0}}\del[p]\mathcal{L}
  &= 2(1-p+\tilde{\eta}_{2}\alpha p)-b^{2}\tilde{\eta}_{2}(\langle n \rangle_{+}-\langle
    n \rangle_{-}) = 0\\[5pt]
  \frac{1}{2f_{0}}\del[\alpha]\mathcal{L}
  &= \tilde{\eta}-2\alpha+\tilde{\eta}_{2}(p^{2}+\alpha^{2})
    +b^{2}\tilde{\eta}_{2}(\langle n \rangle_{+}+\langle n \rangle_{-})=0 
  \end{aligned}
\end{equation}
where we defined the descendant averages
\begin{equation}
  \label{eq:122}
  \langle n \rangle_{\pm}\equiv \langle n\rangle(2\pi
  i\tau_{\pm}(p,\alpha,\tilde{\eta}_{2})),\quad \langle f(n)\rangle(x)
  \equiv \sum_{n}
  \frac{f(n)e^{nx}}{1-e^{nx}},
\end{equation}
with $f(n)$ some arbitrary function of $n$.
We can solve the system \eqref{eq:113} perturbatively by Taylor
expanding $p$, $\alpha$ and the descendant averages in
$\tilde{\eta}_{2}$
\begin{equation}
  \label{eq:123}
  \begin{aligned}
    p &= p_{0} + p_{1}\tilde{\eta}_{2}+\cdots,\\[5pt]
    \alpha &= \alpha_{0} + \alpha_{1}\tilde{\eta}_{2}+\cdots,\\[5pt]
    \langle n \rangle_{\pm} &= \langle n\rangle_{0} -2\pi i \tau
    _{0}(\alpha_{0}\pm p_{0}) \tilde{\eta}_{2}
    \langle n \rangle_{1} +\cdots
  \end{aligned}
\end{equation}
with
\begin{equation}
  \label{eq:131}
  \langle n\rangle_{k} = \del[x]^{k}\langle n\rangle\big\rvert_{x=2\pi
  i\tau_{0}}, \quad k\in \mathbb{N},
\end{equation}
being the connected correlators of $n$. Solving
\eqref{eq:113}, we find the saddle-point values
\begin{equation}
  \label{eq:125}
  \begin{aligned}
    p_{*}(b) &= p_{*}(0) + b^{2}\tilde{\eta}_{2}^{2} [\langle
    n\rangle_{0}- \frac{2\pi^{2}\ell}{\beta}\langle n \rangle_{1}]+\mathcal{O}(\tilde{\eta}_{2}^{3})\\[5pt]
    \alpha_{*}(b) &=\alpha_{*}(0)+b^{2}[\tilde{\eta}_{2}\langle
    n\rangle_{0}+\frac{1}{2}\tilde{\eta}\tilde{\eta}_{2}^{2}(\langle
    n\rangle_{0}+\frac{2\pi^{2}\ell}{\beta}\langle n \rangle_{1} )]
    +\mathcal{O}(\tilde{\eta}_{2}^{3}),
  \end{aligned}
\end{equation}
where $(p_{*}(0),\alpha_{*}(0))$ are the classical values given in
\eqref{eq:104} with $\tilde{\eta}_{3}=0$. The total free energy is then
\begin{equation}
  \label{eq:126}
  \begin{multlined}
    F(b) =2F(0) -2\sum_{n=1}^{\infty}\log(1-e^{-\frac{2\pi^{2}\ell }{\beta}n})
    +\\[5pt]
    +\frac{2\pi^{2}\ell}{\beta}
    \left\{\tilde{\eta}\langle
      n\rangle_{0}\,\tilde{\eta}_{2}+ \left[
        \left(
        \langle
        n\rangle_{0}+ \frac{2\pi^{2}\ell}{\beta}\langle
        n\rangle_{1}\right)
      \left(
        1+\frac{\tilde{\eta}^{2}}{4}
      \right)
        +b^{2}\langle n\rangle_{0}^{2} \right]\tilde{\eta}_{2}^{2}
    \right\}
   +\cdots.
  \end{multlined}
\end{equation}
where $F(0)$ is the classical value in \eqref{eq:97} with $\tilde{\eta}_{3}=0$.

When $\tilde{\eta}_{2}$ is purely imaginary, the real eigenvalue is
\eqref{eq:10}, and we can also do a  finite $c$ saddle-point
computation. In this particular case, the descendant sum does not
depend on real $p$ and $\alpha$ and it is given by
\begin{multline}
  \label{eq:127}
   \sum_{(\lambda,\mu)\in
               \mathbb{Y}_{|\lambda|+|\mu|}}
    e^{-\frac{2\beta}{\ell}(|\lambda|+|\mu|)-\frac{\eta_{2}}{\ell^{2}}
    \left(
      b^{-1}\sum_{j}(\lambda_{j}^{2}+\mu_{j}^{2})+
      b\sum_{j}({\lambda'_{j}}^{2}+{\mu' _{j}}^{2})
               \right)}=\\[5pt]
  =
  \left[
    \sum_{\{\tilde{N}_{m}\}}e^{2\pi i\tau_{0}
  \left(
    \sum_{m}a_{m}\tilde{N}_{m} + \frac{b^{2}\tilde{\eta}_{2}}{2}
    \sum_{n,m}\tilde{N}_{n}A_{nm}\tilde{N}_{m},
  \right)}
  \right]^{2},\quad a_{m} = m+\frac{\tilde{\eta}_{2}}{2}m^{2},
\end{multline}
where we used the free boson representation identities derived in
Appendix \ref{sec:comb-ident} in the second line with the
lower-triangular matrix $A_{nm}$ given by
\eqref{eq:lower-triangular-matrix}. The matrix $A$ is diagonalizable
via $P^{t}\cdot A\cdot P=D$ with $D_{mn}=m\delta_{mn}$, as all the
eigenvalues are distinct. Defining $\tilde{N} = P\cdot \tilde{N}'$ and
$a' = a\cdot P$, we can thus simplify the expression inside brackets
of \eqref{eq:127} to
\begin{equation}
  \label{eq:128}
  \prod_{m=1}^{\infty}\sum_{\tilde{N}'_{m}=0}^{\infty}e^{2\pi i\tau_{0}
    \left(
      a'_{m}\tilde{N'}_{m}+b^{2}\frac{\tilde{\eta}_{2}}{2} m \tilde{N'}_{m}^{2}
    \right) },
\end{equation}
which is an infinite product of \emph{partial theta functions}
\cite{andrews2009ramanujan}, as the sum goes from zero to infinity
instead over all integers.  Partial theta functions have also been
obtained from characters of W-algebras, in connection to false theta
functions and, more generally, to quantum modular forms
\cite{bringmann2015}. Although less studied than the usual theta
functions, partial theta functions still have interesting
quasi-modular properties and identities useful to study its
asymptotics \cite{Bringmann:2017efd,bringmann2019framework}.  We
notice that the classical case $b=0$ of \eqref{eq:128} has a
descendant contribution given by
\begin{equation}
  \label{eq:130}
  \prod_{m=1}^{\infty}\frac{1}{1-e^{2\pi i\tau_{0}(m+\frac{\tilde{\eta}_{2}}{2}m^{2})}}.
\end{equation}
The quadratic power of $m$ above is reminiscent of the Lifshitz free
boson partition function discussed in \cite{Melnikov:2018fhb}, a
non-relativistic correction to the CFT linear dispersion, which also
has non-trivial quasi-modular properties. This partition function will
be explored in more detail in a forthcoming work in connection to
anisotropic field theories and non-relativistic CFTs.

\section{Generalized Eigenstate Thermalization for the qBO$_2$
  charges in the Semiclassical Limit}
\label{sec:geth-qbo_2-charges}

In the previous section, we saw that the integrable structure of the
qBO$_{2}$ hierarchy allow us to describe a great deal of its GGE,
including quantum corrections. Now we want to explore the Generalized
ETH for the qBO$_{2}$ charges, which is interpreted here as a
comparison between matrix elements and ensemble averages of these
charges.

First, let us briefly review the usual justification for why the ETH
is equivalent to thermalization of a quantum many-body system. Given a
Hamiltonian $H$ with no charged interactions, its orthonormal
eigenbasis $|E_{n},q\rangle$ with a commuting charge $q$, and an
initial state $|\psi_{0}\rangle$, consider the time-evolved quantum
state
\begin{equation}
  \label{eq:time_dep_wave_fct}
  |\psi(t)\rangle = e^{-iHt}|\psi_{0}\rangle =
  \sum_{n,q}c_{n,q} e^{-iE_{n}t}|E_{n},q\rangle,\quad c_{n,q} = \langle E_{n},q|\psi_{0}\rangle.
\end{equation}
The expectation value of an observable $\mathcal{O}$, which by
assumption does not couple sectors of different charges, is given by
\begin{multline}
  \label{eq:time_dep_obs_avg}
  \langle \psi(t)| \mathcal{O} |\psi(t)\rangle =
  \sum_{\substack{n,m,q \\ E_{n} \neq
  E_{m}}}e^{-i(E_{n}-E_{m})t}c_{n,q}c^{*}_{m,q}
  \langle E_{m},q|\mathcal{O}|E_{n},q\rangle+
  \\[5pt]+\sum_{\substack{n,m,q \\ E_{n} = E_{m}}}c_{n,q}c^{*}_{m,q} \langle E_{m},q|\mathcal{O}|E_{n},q\rangle.
\end{multline}
Taking a time average of \eqref{eq:time_dep_obs_avg} over a period
$\tau$, denoted by $\langle\, .\, \rangle_{\tau}$, and assuming there
is no degeneracy in the spectrum, we get
\begin{equation}
  \label{eq:diagonal_ensemble}
  \langle\langle \psi(t)| \mathcal{O}
  |\psi(t)\rangle \rangle_{\tau}
  = \sum_{n,q}|c_{n,q}|^{2} \langle E_{n},q|\mathcal{O}|E_{n},q\rangle,
\end{equation}
the so-called \emph{diagonal ensemble}. This ensemble strongly depends
on the initial state via the coefficients $c_{n}$, and thus there is
no apparent thermalization. The ETH is an assumption over the
statistical distribution of matrix elements over the energy ensemble,
making the diagonal ensemble equivalent to the microcanonical one in
the thermodynamic limit. Thus, the many-body ETH can be stated as
\begin{equation}
  \label{eq:eth-2}
  \langle E_{n}, q|\mathcal{O}|E_{m},q\rangle
  = \mathcal{O}_{\text{th}}(\bar{E},q)\delta_{nm}
  + e^{-S(\bar{E},q)}f_{\mathcal{O}}(E_{n},E_{m},q)R_{nm},
\end{equation}
where $\bar{E} \equiv (E_{n}+E_{m})/2$, $\mathcal{O}_{\text{th}}$ is a
thermal average, $S$ the entropy, $f_{\mathcal{O}}$ is a slowly
varying function and $R_{nm}$ is a random matrix of zero mean and unit
variance \cite{Deutsch2018,Das:2017vej,Belin:2020jxr}. The equivalence
between the ensembles is strictly true in the thermodynamic limit,
given that $S$ is an extensive quantity. The ETH might seem artificial
in this presentation, but there are strong arguments that it should be
valid for chaotic, non-integrable isolated quantum systems.

A more refined formulation of ETH is to consider the initial states as
a result of a non-integrable, random perturbation over unperturbed
energy eigenstates. Under certain reasonable assumptions, one can then
derive the ETH form \eqref{eq:eth-2} \cite{Deutsch2018}. An
alternative formulation of \eqref{eq:eth-2} in the thermodynamic limit
is that the diagonal matrix elements are given by smooth functions of the
energy and charge \cite{Lashkari:2016vgj}
\begin{equation}
  \label{eq:119}
   \langle E_{n},q|\mathcal{O}|E_{n},q\rangle = \mathcal{O}_{\text{eth}}(E_{n},q).
\end{equation}
For an integrable system, the eigenstates can depend on an infinite
set of conserved charges $Q_{k}$. In the CFT case, all the higher
charges can be written in terms of the energy and U(1)
charge. Therefore, the Generalized ETH in the thermodynamic limit can
be formulated as \cite{Dymarsky:2019etq}
\begin{equation}
\label{eq:120}
\langle E_{n},q|\mathcal{O}|E_{n},q\rangle = \mathcal{O}_{\text{geth}}(Q_{k}(E_{n})).
\end{equation}
A fundamental question of thermalization is to answer if the proposed
smooth functions are equivalent to an ensemble average. Here, we will
compare $\mathcal{O}_{\text{geth}}$ with the related GGE ensemble
average for the qBO$_{2}$ charges. As discussed in the introduction,
while this analysis might be relevant to verify the ensemble
equivalence for operators in the vacuum family of holographic CFTs, it
does not address the stronger version of generalized ETH for generic
operators in 2D CFTs, which remains an open problem. Therefore, every
mention on ETH below should be interpreted in the context of ensemble
equivalence of conserved charges and operators in the vacuum family.

In this work, we are interested in the ETH for charged 
CFTs.  Previous works noticed that the standard ETH is valid only for
\emph{primary states} in the strict large central charge limit
\cite{Basu_2017,Lashkari:2016vgj}, with $1/c$ corrections breaking the
equality between ensembles. 
In another development, Calabrese and Cardy studied perturbations of
2D CFTs under a quantum quench \cite{Calabrese_2007}, concluding that
the now called Calabrese-Cardy states thermalize to the usual Gibbs
ensemble. Later Cardy studied more generic states, perturbed by
irrelevant deformations built from powers and derivatives of the
energy-momentum tensor, and concluded that those thermalize to a
\emph{non-abelian} GGE, with conserved charges not necessarily
commuting among themselves \cite{Cardy_2016}. The qKdV GGE is formed
by taking only a commuting subset of these more generic conserved
charges \cite{Bazhanov:1994ft}. The ETH failure for descendants and
under the inclusion of $1/c$ corrections, added to the observations
that perturbed CFT states depend on higher conserved charges, justify
the investigation of GGEs in 2D CFTs. The qKdV GGE was studied in
\cite{Dymarsky:2018lhf,Dymarsky:2018iwx,Maloney:2018yrz,Maloney:2018hdg}
and the Generalized ETH for this case was proposed in
\cite{Dymarsky:2019etq}. There are some subtleties to be considered in
the charged ETH when sectors with different charges interact
\cite{Belin:2020jxr}. However, here we are interested in the
Generalized ETH for the subset of qBO$_{2}$ charges in U(1) charged 2D
CFTs. Although those are U(1) charged operators, they are diagonal in
the AFLT basis and do not have interaction terms coupling distinct
charged sectors. Therefore, we can consider the standard ETH condition
as formulated above for the qBO$_{2}$ charges.

Let $|P,a\rangle$ be a normalized primary eigenstate. The diagonal
elements of the charges are
\begin{equation}
  \label{eq:99}
   \langle P,a| Q_{k}|P,a\rangle
 = \frac{E^{(k)}(P,a)}{\ell^{k}}
\end{equation}
and the strict Generalized ETH is valid if, in the thermodynamic limit,
\begin{equation}
  \label{eq:103}
  \langle Q_{k}\rangle_{GGE} = \del[\eta_{k}]F =\frac{E^{(k)}(P,a)}{\ell^{k}}.
\end{equation}
This comparison is a first step to analyze the Generalized ETH for the
vacuum Verma module, relevant to the AdS$_{3}$/CFT$_{2}$
correspondence \cite{Dymarsky:2019etq}. This requires a more detailed
study of the structure of quasi-primary operators for charged CFTs, to
be investigated in a future work.

Although we do not have the GGE free energy in closed form for finite
$c$, we can still check if the Generalized ETH for the charges can be
satisfied in the thermodynamic semiclassical limit using the results
of the previous section. We want to fix the chemical potentials to
make eq.~\eqref{eq:103} true for any values of $P$ and $a$.  To
leading order in the semiclassical limit, the free energy is exactly
given by its the saddle-point value
\begin{equation}
  \label{eq:101}
  F(\bm{\eta}) = \mathcal{L}(P_{*},a_{*}) =
  2\pi\sqrt{\frac{c\Delta_{*}}{6}}
+\sum_{k\geq 0}\frac{\eta_{k}}{\ell^{k}}{E^{(k)}(P_{*},a_{*})}.
\end{equation}
This implies that the GETH condition \eqref{eq:103} is automatically
satisfied in the thermodynamic, semiclassical limit if
$P = P_{*}(\bm{\tilde{\eta}})$ and $a =
a_{*}(\bm{\tilde{\eta}})$. Given that the saddle-point values are
functions of the chemical potentials, we are fixing two of them to
match the arbitrary $P$ and $a$ (typically the inverse temperature
$\beta$ and the chemical potential $\eta$).  All the other charges
automatically match as they are polynomials of $P$ and $a$. This is
equivalent to what happens in the qKdV hierarchy
\cite{Dymarsky:2018lhf,Maloney:2018yrz}.

Now let us consider the leading quantum correction to  the free
energy  in \eqref{eq:118}. It is tempting to make the
same choice as before, $P=P_{*}, a=a_{*}$, such that
\begin{align}
  \label{eq:117}
  \langle Q_{k}\rangle_{GGE} - \frac{E^{(k)}(P,a)}{\ell^{k}}
  &\sim  
    \frac{13\pi^{2}\ell}{6}\del[\eta_k](\frac{p_{*}}{\beta})
    -\sum_{j=1,2}\sum_{m>0}\del[\eta_k]\log
    \left(
    1-e^{-
    \tilde{g}_{j,m}}
    \right)
    \nonumber\\[5pt]
  &\approx 
    \frac{13\pi^{2}\ell}{6}\del[\eta_k](\frac{p_{*}}{\beta})
    -\sum_{j=1,2}\int_{0}^{\infty} dm\,\frac{\del[\eta_k]\tilde{g}_{j,m}}
    {e^{\tilde{g}_{j,m}}-1},
\end{align}
where in the second step we replaced the sum over $m$ by an integral,
which is a good approximation in the thermodynamic limit.  Two
chemical potentials have been fixed to make the zero order value of
the charges to match and all the charges have a distinct subleading
correction given by \eqref{eq:117}. Therefore, even if it is possible
to separately cancel the corrections in principle, there is not enough
chemical potentials to set all the subleading corrections to
zero. This implies in a failure to match the GGE averages with primary
state averages.  On the other hand, we do not consider this a proof
that the Generalized ETH does not work at $1/c$ order, as the choice
$P=P_{*}, a=a_{*}$ seems to be a particular
ansatz.
To achieve a concrete proof of this matter requires a better
understanding of the quantum corrections and of the analytic structure
of the saddle-point. Thus, we leave the analysis of the validity of
the charged Generalized ETH including quantum corrections for the
future.

\section{Discussion}
\label{sec:conclusions}

In this work, we presented the first steps to calculate the qBO$_{2}$
GGE averages for charged 2D CFTs using the AGT correspondence. While
our setup was rather general, our concrete calculations focused in the
case in which only the first two non-trivial charges $Q_{2}$ and
$Q_{3}$ were non-zero. The crucial ingredient to compute the GGE
partition function was the AFTL basis, an alternative basis for
descendants of the Virasoro and U(1) algebras, which diagonalizes all
the qBO$_{2}$ charges. The associated eigenvalues are generically
complex, being a caveat to the equilibrium ensemble interpretation for
chiral CFTs. However, those can make sense for non-chiral CFTs if we
sum the holomorphic and anti-holomorphic charges, giving a real total
contribution. While writing the qBO$_{2}$ GGE partition function, we
argued that the trace is equivalent to the case in which the Virasoro
and U(1) modes do not commute, being the usual convention in the
literature.

From the qBO$_{2}$ GGE partition, we obtained the free energy in the
thermodynamic semiclassical limit, including the leading quantum
correction in the Liouville parameter $b$, equivalent to the leading
correction in the central charge. Focusing on the first two
non-trivial eigenvalues, we obtained the perturbative saddle-point
solutions and discussed the conditions for the partition function
convergence and boundedness of the free energy. We also presented the
saddle-point analysis at finite $c$ with only $\eta_{2}$ being
non-zero. For $\eta_{2}$ being real, we obtained a perturbative
expansion of the free energy depending on descendant averages.  For
purely imaginary $\eta_{2}$, we obtained the exact descendant sums in
the partition function in terms of partial theta functions. This
example entails the main technical point of this paper, an alternative
approach to obtain finite $c$ GGE averages using the AFLT
basis.

In the last part of this paper, we briefly discussed the Generalized
ETH for the qBO$_{2}$ charges in the semiclassical limit, which is
definitely obeyed at leading order in the central charge, just like
the qKdV case. Whilst we argued that the Generalized ETH is not valid
when $1/c$ corrections are included, we deemed the argument
preliminary as it is unclear how much it relies on the particular
ansatz we made. We shall thus leave the appropriate analysis of the
Generalized ETH of charged CFTs for a future work. For recent results
on the ETH for charged CFTs, see \cite{Das:2017vej,Belin:2020jxr}. We
emphasize that the analysis presented here should be relevant to the
generalized ETH for the vacuum family in the semiclassical limit, but
it should have limited application in the generalized ETH for generic
heavy operators, whose averaged version without extra conserved
charges is discussed in \cite{Collier:2019weq,Das:2020uax}.

There are many interesting forthcoming directions to our work. First,
our analysis could be adapted to the qILW$_{2}$ hierarchy, whose
spectrum has been conjectured in \cite{Litvinov:2013zda} and
interpolates between the qBO$_{2}$ and the qKdV plus free boson
one. Therefore, although technically challenging, the exact qKdV
spectrum and GGE partition function can be obtained by taking the
appropriate limit of the qILW$_{2}$ results. Another interesting
direction is to obtain the qKdV GGE via the ODE/IQFT correspondence,
as suggested in \cite{Kotousov:2019ygw}, which obtained the first few
qKdV eigenvectors, the analogous of the AFLT basis in this case.  It
would also be intriguing to study the GGE with extra parafermionic or
non-local charges, as suggested by \cite{Cardy_2016,Kotousov:2019nvt}.
The need to add quasi-local charges in the generalized ETH was
observed in lattice integrable systems \cite{Ilievski:2015aa}, leaving
open the possibility that these charges might improve matters at
finite $c$. Finally, as pointed out in our finite $c$ free energy
computation, partial theta functions appear, which have modular
properties related to false theta functions and quantum modular forms
\cite{bringmann2015,Bringmann:2017efd,bringmann2019framework}. As we
pointed out, these can be relevant in the studies of anisotropic field
theories and non-relativistic CFTs \cite{Melnikov:2018fhb} (see also
\cite{Fuentealba:2019oty}). In another development,
\cite{Maloney:2018hdg} showed that correlation functions of qKdV
charges have modular properties, obeying certain modular differential
equations. It would also be worthwhile to investigate if the qBO$_{2}$
correlation functions obey similar modular equations.

The natural holographic realization of the qBO$_{2}$ hierarchy is
AdS$_{3}$ gravity with an extra U(1) gauge field, as described in
\cite{Melnikov:2018fhb}, while the chiral case with only one
SL(2)$\times$U(1) current should be related to Warped CFTs (WCFT)
\cite{Detournay:2012pc,Hofman:2014loa,Azeyanagi:2018har}, initially
proposed as holographic duals of near-extremal black holes.  We have
seen that the qBO$_{2}$ spectrum is not straightforwardly adapted to
chiral CFTs, due to its complex eigenvalues, requiring further studies
in the particular case of WCFTs and its related black hole solutions,
in line with recent discussions in the KdV case
\cite{Perez:2016vqo,Dymarsky:2020tjh}.

The results presented here can also be useful to explore
thermalization in effective field theory descriptions of lattice
systems, in which higher integrable charges can be interpreted as
precise deformations of the 2D CFT description \cite{lukyanov1998low}.
Along these lines, the quantum Benjamin-Ono$_{2}$ equation has been
proposed to describe non-linear deviations from the CFT description in
the fractional Quantum Hall liquid
\cite{abanov2005quantum,bettelheim2006quantum,bettelheim2007nonlinear,Abanov:2008ft},
suggesting a concrete physical system to apply our results.

\section*{Acknowledgments}
\label{sec:acknowledgments}

I would like to thank stimulating discussions with Marcone Sena, David
Tempo and Ricardo Troncoso about this work. The Centro de Estudios
Científicos (CECs) is funded by the Chilean Government through the
Centers of Excellence Base Financing Program of Conicyt.

\appendix{}

\section{Derivation of Integrals of Motion and Analytic Ordering}
\label{sec:deriv-integr-moti}

Given two local operators $A(z)$ and $B(w)$, their operator product
expansion have divergent terms that need to be regulated. The standard
procedure to regularize this product in CFT is called \emph{analytic
  ordering}, a generalization of the normal ordering for free fields
\cite{Bazhanov:1994ft,DiFrancesco1997a}. An algorithm to calculate the
analytic ordering on the cylinder has been pedagogically described in
\cite{Dymarsky:2019iny}. Here we present a variation of that approach
using OPEs and including $U(1)$ terms with Hilbert transforms.

The main idea of analytic ordering is to make a Cauchy integral that
removes all the singular parts in the coincidence limit
$z\rightarrow w$. The analytic ordering of two operators $A(z)$ and
$B(z)$ on the cylinder is then defined as
\begin{equation}
  \label{eq:analytic_order}
  (AB)(w) = \cauchy{} \frac{1}{z-w}\mathcal{T}(A(z)B(w)),
\end{equation}
where
\begin{equation}
  \label{eq:time_ordering}
  \mathcal{T}(A(z)B(w)) =\left\{
  \begin{split}
    A(z)B(w),\quad \Re z < \Re w,\\[5pt]
    B(w)A(z),\quad \Re z > \Re w,
  \end{split}\right.
\end{equation}
is the time-ordering operator. This definition is the same for
operators defined on the plane, but instead of time-ordering (real
part of $z$), we use radial ordering. When expressed in terms of the
$A$ and $B$ modes on the plane, the two ordering procedures
differ. Our strategy here is to rewrite \eqref{eq:analytic_order} in
terms of coordinates and operators on the plane. In the following, we
simply differ operators on the cylinder and on the plane by its
coordinates: $A(z), B(w)$ for the former and $A(u), B(v)$ for the
latter.

Any OPE can be expressed as a singular part and a regular part
\cite{DiFrancesco1997a}
\begin{equation}
  \label{eq:ope_decomposition}
  A(u)B(v) = \overbracket{A(u)B}(v)+ (A(u)B(v)),
\end{equation}
where the singular part is
\begin{equation}
  \label{eq:singular_ope}
  A(u)B(v) \,\sim\, \overbracket{A(u)B}(v) = \sum_{k>0}\frac{\{AB\}_{k}(v)}{(u-v)^{k}}
\end{equation}
and the regular part
\begin{equation}
  \label{eq:regular_ope}
  (A(u)B(v)) = \sum_{k\geq 0} \frac{(u-v)^{k}}{k!}(\del[]^{k}A B)(v). 
\end{equation}
The  $\sim$ sign represents the display of only the singular
part.  If $A, B$ have holomorphic weights $h_{A},h_{B}$ respectively,
we expand the operators on the plane as
\begin{equation}
  \label{eq:plane_expansion}
  A(u) = \sum_{n}A_{n}u^{-n-h_{A}},\quad    B(u) = \sum_{n}B_{n}u^{-n-h_{B}}.
\end{equation}
If we define 
\begin{equation}
  \label{eq:ope_expansion}
  (AB)(u) = \sum_{n}(AB)_{n}\,u^{-n-h_{A}-h_{B}},
\end{equation}
using the analytic ordering on the plane, we get
\cite{DiFrancesco1997a}
\begin{equation}
  \label{eq:ope_modes}
  (AB)_{m} = \sum_{n\leq -h_{A}} A_{n}B_{m-n} + \sum_{n > -h_{A}} B_{m-n}A_{n}.
\end{equation}

We can map the cylinder to the plane by a
conformal transformation $u= e^{z}$. In general, primary fields
transform as
\begin{equation}
  \label{eq:6}
  A(z) = 
  \left(
    \frac{dz}{du}
  \right)^{-h_{A}}A(u) = u^{h_{A}}A(u) \equiv \tilde{A}(u), 
\end{equation}
where we remind that we distinguish the operator domains by their
argument. Operators on the cylinder have a mode expansion around $u=0$
equivalent to an operator of zero weight on the plane
\begin{equation}
  \label{eq:36}
  \tilde{A}(u) = \sum_{n}\tilde{A}_{n}u^{-n}.
\end{equation}
This is also the case of the energy-momentum tensor $T(z)$, even
though it is a quasiprimary. We want to express the integral
\eqref{eq:analytic_order} in plane coordinates
\begin{align}
  \label{eq:integrand-1}
  \frac{dz}{2\pi i} \frac{u^{h_{A}}A(u)v^{h_{B}}B(v)}{z-w}
  = \frac{du}{2\pi i u}\,
    \frac{1}{\ln(1+\frac{u-v}{v})}\;\tilde{A}(u)\tilde{B}(v).
\end{align} 
The integrand can be rewritten in terms of the \emph{Bernoulli polynomials
of the second kind} via its generating function \cite{jordan1965calculus}
\begin{equation}
\label{eq:gen-fct-bernoulli-2nd-kind}
  \frac{(z+1)^{x}}{\ln(1+z)} =
  \sum_{n=0}^{\infty}\psi_{n}(x)z^{n-1},\quad |z| < 1.
\end{equation}
The first five
polynomials are
\begin{equation}
  \begin{gathered}
    \psi_{0}(x) = 1,\quad \psi_{1}(x) = x + \frac{1}{2},\quad
    \psi_{2}(x) = \frac{1}{2}x^{2}-\frac{1}{12},\\[5pt]
    \psi_{3}(x) = \frac{1}{6}x^{3}-\frac{1}{2}x^{2}+\frac{1}{24},\quad
    \psi_{4}(x) = \frac{1}{24}x^{4} -
    \frac{1}{6}x^{3}+\frac{1}{6}x^{2}-\frac{19}{720}.
  \end{gathered}
\end{equation}
The simplest relevant values for us are $x=0$ and $x=1$
\begin{equation}
  \label{eq:bernoulli-simplest}
  \psi_{n}(0) = G_{n},\quad \psi_{n}(1) = G_{n-1} + G_{n},
\end{equation}
where $G_{n}$ are the reciprocal logarithmic numbers or Gregory
coefficients\footnote{These can be also easily obtained from
  $G_{n} = -\frac{B_{n}^{(n-1)}(0)}{(n-1)n!}$, where $B_{n}^{(s)}(z)$
  are the Generalized Bernoulli polynomials or Norlund polynomials,
  implemented in Mathematica as \texttt{NorlundB[n,s,z]}. } obtained
from
\begin{equation}
  \label{eq:gen-fct-gregory-coeffs}
\frac{z}{\log(1+z)} = 1+\frac{1}{2}z -\frac{1}{12}z^{2} +
\frac{1}{24}z^{3}-\frac{19}{720}z^{4}+\dots = 1 + \sum_{n=1}^{\infty}G_{n}z^{n}.
\end{equation}
Using \eqref{eq:gen-fct-bernoulli-2nd-kind} with $z = \frac{u-v}{v}$
and $x=0$ in \eqref{eq:integrand-1}, we get from
\eqref{eq:analytic_order}
\begin{align}
  \label{eq:cylinder-to-plane-ordering}
  (AB)(w)
  &=\oint_{v}\frac{du}{2\pi i u}
    \left(
    \frac{v}{u-v} + \sum_{n=1}^{\infty}\psi_{n}(0)
    \left(
    \frac{u-v}{v}
    \right)^{n-1}
    \right)\;\tilde{A}(u)\tilde{B}(v)\nonumber\\[5pt]
  &= (\tilde{A}_{-}\tilde{B}_{+})(v)
    + \cauchy[u]{v}
    \sum_{n=1}^{\infty}\psi_{n}(0)
    \left(\frac{u-v}{v}\right)^{n-1}
    \; u^{h_{A}-1}v^{h_{B}} \sum_{k=1}^{h_{A}+h_{B}}
    \frac{ \{AB\}_{k}(v)}{(u-v)^{k}}.
\end{align}
The first term above is the analytic ordering on the plane of the
operators defined as
\begin{equation}
  \label{eq:44}
  \tilde{A}_{\pm}(u) = \sum_{n}\tilde{A}_{n}u^{-n\pm 1}.
\end{equation}
The modes of $(\tilde{A}_{-}\tilde{B}_{+})(v)$ are given by
\eqref{eq:ope_modes} with $h_{\tilde{A}_{-}}=1$
\begin{align}
  \label{eq:59}
  (\tilde{A}_{-}\tilde{B}_{+})_{m}
  &= \sum_{n\leq -1}
    \tilde{A}_{n}\tilde{B}_{m-n} +
    \sum_{n > -1}\tilde{B}_{m-n}\tilde{A}_{n}\nonumber \\[5pt]
  &= \sum_{n\geq 1}
    (\tilde{A}_{-n}\tilde{B}_{m+n} + \tilde{B}_{m-n}\tilde{A}_{n}) + \tilde{B}_{m}\tilde{A}_{0}.
\end{align} 
The second term depends only on the singular OPE, which
can be computed with the methods described in
\cite{DiFrancesco1997a}. As it stands, the OPE will produce
derivatives of the function $u^{h_{A}-1}$. Assuming $h_{A} \geq 1$, we
can Taylor expand this function around $u=v$
\begin{align*}
  u^{h_{A}-1} = v^{h_{A}-1}\sum_{p=0}^{h_{A}-1}
  \begin{pmatrix}
    h_{A}-1 \\ p
  \end{pmatrix} 
  \left(\frac{u-v}{v}\right)^{p},
\end{align*}
and then
\begin{align*}
  &v^{h_{A}+h_{B}-1}\oint_{v}\frac{du}{2\pi i}
    \sum_{k=1}^{h_{A}+h_{B}}\sum_{n=1}^{\infty}\sum_{p=0}^{h_{A}-1}
    \begin{pmatrix}
      h_{A}-1 \\ p
    \end{pmatrix}
  \psi_{n}(0)
  \left( u-v\right)^{n+p-k-1}
  \; 
  v^{1-n-p}\{AB\}_{k}(v)\\[5pt]
  &= 
    \sum_{k=1}^{h_{A}+h_{B}} \sum_{n=1}^{\infty}\sum_{p=0}^{h_{A}-1}
    \begin{pmatrix}
      h_{A}-1 \\ p
    \end{pmatrix}
  \psi_{n}(0)\delta_{n+p,k}
  \; 
  v^{h_{A}+h_{B}-k}\{AB\}_{k}(v)\\[5pt]
  &= 
    \sum_{k=1}^{h_{A}+h_{B}} 
    \left(
    \sum_{n=1}^{k}
    \begin{pmatrix}
      h_{A}-1 \\ k-n
    \end{pmatrix}
  \psi_{n}(0)
   \right)
  \; 
  v^{h_{A}+h_{B}-k}\{AB\}_{k}(v)\\[5pt]
 &= 
    \sum_{k=1}^{h_{A}+h_{B}} 
   \left(\psi_{k}(h_{A}-1) -
    \begin{pmatrix}
      h_{A}-1 \\ k
    \end{pmatrix}\right)
  \; 
  v^{h_{A}+h_{B}-k}\{AB\}_{k}(v),
\end{align*}
where in the fourth line we used the identity
\begin{equation}
  \label{eq:psi-identity}
  \sum_{n=1}^{k}
    \begin{pmatrix}
      h_{A}-1 \\ k-n
    \end{pmatrix}
    \psi_{n}(0) = \psi_{k}(h_{A}-1) -
   \binom{h_{A}-1}{k}.
\end{equation}
If we expand the OPE coefficients as
\begin{equation}
  \label{eq:regulated-ope-expansion}
  \{AB\}_{k}(v) = \sum_{m}\{AB\}_{k,m}\;v^{-h_{A}-h_{B}+k-m},
\end{equation}
going back to \eqref{eq:cylinder-to-plane-ordering}, we get
\begin{align}
  \label{eq:48}
  (AB)(w) = \sum_{m}[ (\tilde{A}_{-}\tilde{B}_{+})_{m}+
 \{AB\}_{m}]v^{-m}
\end{align}
with 
\begin{align}
  \label{eq:49}
  \{AB\}_{m} \equiv     \sum_{k=1}^{h_{A}+h_{B}} 
 \left( \psi_{k}(h_{A}-1) -
    \binom{h_{A}-1 }{k}
 \right) \{AB\}_{k,m}.
\end{align}
The OPE modes can be obtained from the inverse formula of
\eqref{eq:regulated-ope-expansion}
\begin{equation}
  \label{eq:regulated-ope-modes}
   \{AB\}_{k,m} = \cauchy[v]{} v^{h_{A}+h_{B}-k+m-1}\{AB\}_{k}(v).
\end{equation}
From this, the zero mode on the
cylinder is given
by
\begin{align}
  \label{eq:zero-mode-cylinder}
  (AB)_{0} &= \oint \frac{dw}{2\pi i} (AB)(w) \nonumber\\[5pt]
  &= (\tilde{A}_{-}\tilde{B}_{+})_{0} + \{AB\}_{0}.
\end{align}
Therefore, we reduced the problem of analytic ordering to finding the
coefficients of the OPEs between operators. We can recursively use the
definitions above to find the qBO$_{2}$ charges. Notice that the
procedure needs to be slightly modified if the operators are
descendants, i.e., involve derivatives of primaries, and for the
energy-momentum tensor. We do not display a general formula for these
cases, but we will show some examples below on how to proceed.

Let us apply the zero mode formula \eqref{eq:zero-mode-cylinder} for
the simplest cases $J^{2}$ and $T^{2}$. The currents $T$ and $J$
change under $u=e^{z}$ as
\begin{equation}
  \label{eq:cylinder_to_plane}
  J(z)= u J(u) = \sum_{n}a_{n}u^{-n},\quad
  T(z)= u^{2}T(u) - \frac{c}{24} =
  \sum_{n}\tilde{L}_{n}u^{-n},
\end{equation}
where
\begin{equation}
  \label{eq:shifted-virasoro}
  \tilde{L}_{n} = L_{n} -\frac{c}{24}\delta_{n,0}.
\end{equation}
Thus, although $T$ is a quasiprimary, we can directly apply the
formulas above for $T$ if we use the shifted modes in the first term
of \eqref{eq:zero-mode-cylinder}. The fundamental singular OPEs of $J$
and $T$ on the plane are
\begin{equation}
  \label{eq:JJ-ope}
  J(u)J(v) \sim \frac{1/2}{(u-v)^{2}},\quad T(u)J(v) \sim 0,
\end{equation}
and
\begin{equation}
  \label{eq:TT-ope}
  T(u)T(v) \sim
  \frac{c/2}{(u-v)^{4}} + \frac{2 T(v)}{(u-v)^{2}} + \frac{\del[]T(v)}{u-v}.
\end{equation}
From \eqref{eq:ope_modes}, we have that
\begin{align}
  \label{eq:38}
  (\tilde{J}_{-}\tilde{J}_{+})_{0} = \sum_{n\leq -1} a_{n}a_{-n} + \sum_{n > -1}
  a_{-n}a_{n}= 2 \sum_{n > 0}
  a_{-n}a_{n} + a_{0}^{2}.
\end{align}
From the definition \eqref{eq:singular_ope}, the $JJ$ OPE
\eqref{eq:JJ-ope} and the Bernoulli polynomials $\psi_{n}(0)$ obtained
from \eqref{eq:bernoulli-simplest} and
\eqref{eq:gen-fct-gregory-coeffs}, we find
\begin{equation}
 \label{eq:35}
 \{\tilde{J}_{-}\tilde{J}_{+}\}_{0} =
 \sum_{k=1}^{2}\psi_{k}(0) \{JJ\}_{k,0} = -\frac{1}{24},
\end{equation}
where we used that $\binom{0}{x} = 0$, $x>0$. The zero mode on the cylinder
is thus the sum of \eqref{eq:38} and \eqref{eq:35}
\begin{equation}
  \label{eq:JJ-zero-mode}
   (JJ)_{0} =  2 \sum_{n > 0}
   a_{-n}a_{n} + a_{0}^{2} -\frac{1}{24}.
\end{equation}
For the TT zero mode, we get
\begin{align}
  \label{eq:TT-zero-mode}
  (TT)_{0}
  &= \sum_{n\leq -1}\tilde{L}_{n}\tilde{L}_{-n} +\sum_{n>-1}\tilde{L}_{-n}\tilde{L}_{n}+  \sum_{k=1}^{4}\psi_{k}(1)
    \{TT\}_{k,0} - \{TT\}_{1,0}\nonumber\\[5pt]
  &=2\sum_{n > 1}L_{n}L_{-n} + \tilde{L}_{0}^{2} +
     \frac{11}{720}\frac{c}{2} + \frac{5}{12}
    2L_{0} - \frac{3}{2} 2L_{0}+2L_{0}\nonumber\\[5pt]
  &= 2 \sum_{n>0}L_{-n}L_{n} + L_{0}^{2} - \frac{c+2}{12}L_{0}
    +\frac{c^{2}}{(24)^{2}}+ \frac{11c}{1440}\nonumber\\[5pt]
  &= 2 \sum_{n>0}L_{-n}L_{n} + L_{0}^{2} - \frac{c+2}{12}L_{0}
    +\frac{5 c^{2}+ 22 c}{2880},
\end{align}
which matches the known result
\cite{Bazhanov:1994ft,Dymarsky:2019iny}.
 
\subsection{Derivation of $Q_2$}
\label{sec:derivation-q_2}

To complete  our minimal set of examples, let us derive $Q_{2}$ in
\eqref{eq:bo2-operators}. The first term  $TJ$ is easy as its singular
OPE is
zero
\begin{align}
  \label{eq:TJ-zero-mode}
  (TJ)_{0} = (\tilde{T}_{-}\tilde{J}_{+})_{0} = \sum_{n}
  \tilde{L}_{-n}a_{n} =
  \left[
   \sum_{n\neq0}L_{-n}a_{n} + a_{0}\left(L_{0}-\frac{c}{24}\right)
  \right].
\end{align}
The third term is
\begin{align}
  \label{eq:J3-zero-mode-1}
  (J^{3})_{0} &= (J(J^{2}))_{0} = 
                \left[
                \sum_{m\leq -1}
                a_{m}(J^{2})_{-m}+\sum_{m > -1}(J^{2})_{-m}a_{m}
                \right]+
                \sum_{k>0}\psi_{k}(0)\left\{J(J^{2})\right\}_{k,0}.
\end{align}
Using that
\begin{equation}
  \label{eq:J2}
  (J^{2})_{m} =
      \sum_{n\leq -1} a_{n}a_{m-n}+\sum_{n> -1} a_{m-n}a_{n}
    -\frac{1}{24}\delta_{m,0},
\end{equation}
the first two terms inside brackets in \eqref{eq:J3-zero-mode-1}
simplify to
\begin{align}
  \label{eq:60}
  &\sum_{m\leq -1} \sum_{n\leq -1} a_{m}a_{n}a_{-m-n} +
    \sum_{m\leq -1} \sum_{n> -1} a_{m}a_{-m-n}a_{n} +
    \sum_{m> -1} \sum_{n \leq -1} a_{n}a_{-m-n}a_{m}
    -\frac{1}{24}a_{0}
    \nonumber\\[5pt]
  &+\sum_{m > -1} \sum_{n > -1} a_{-m-n}a_{n}a_{m} =
    \sum_{n+m+p=0} :a_{n}a_{m}a_{p}:  -\frac{1}{24}a_{0},
\end{align}
where $::$ denotes the usual normal ordering of placing the highest
modes to the right. To obtain the singular OPE, we use the generalized Wick
theorem \cite{DiFrancesco1997a}
\begin{align}
  \label{eq:47}
  \overbracket{J(z)(J}J)(w)= \cauchy[x]{} \frac{1}{x-w}
  \left(
  \overbracket{J(z)J}(x)J(w) + J(x)\overbracket{J(z)J}(w) 
  \right) = \frac{J(w)}{(z-w)^{2}},
\end{align}
One can  show by induction that
\begin{equation}
  \label{eq:62}
  \overbracket{J(z)(J^{n}})(w) = \frac{n}{2}\frac{(J^{n-1})(w)}{(z-w)^{2}},
\end{equation}
which then implies that
\begin{equation}
  \label{eq:63}
  \left\{J(J^{n})\right\}_{k,m} = \frac{n}{2} (J^{n-1})_{m}\,\delta_{k,2}.
\end{equation}
Going back to \eqref{eq:J3-zero-mode-1}, we get
\begin{equation}
  \label{eq:J3-zero-mode}
  \begin{aligned}
    (J^{3})_{0} &= \sum_{n+m+p=0} :a_{n}a_{m}a_{p}: -\frac{1}{24}a_{0}
    + \psi_{2}(0)a_{0}\\[5pt]
    &= \sum_{\substack{n+m+p=0 \\ n,m,p\neq 0}} :a_{n}a_{m}a_{p}: +
    a_{0}^{3}+3a_{0} \left( 2\sum_{n>0}a_{-n}a_{n} -\frac{1}{24}
    \right)
  \end{aligned}
\end{equation}

The second term in $Q_{2}$ is $(J\mathcal{H}\del[x]J)$ and requires us to deal
with a derivative and the Hilbert transform
\eqref{eq:hilbert-transform-circle}. Let us define the Hilbert
transform derivative operator \begin{equation}
  \label{eq:64}
 \hilbert \equiv    i\mathcal{H}u\del[u] = \mathcal{H}\del[x] ,
\end{equation}
where we used that $u = e^{ix}$. The analytic continuation of the Hilbert transform on the circle
\eqref{eq:hilbert-transform-circle} can be written as
\begin{equation}
  \label{eq:hilbert-transform}
  \mathcal{H}f(u) = i\mathcal{P}\oint_{C}
  \frac{d\zeta}{2\pi i} f(\zeta) \frac{\zeta+u}{\zeta(\zeta-u)},
\end{equation}
where the contour $C$ passes through $\zeta = u$ (which is why we need
the principal value integral). To calculate
\eqref{eq:hilbert-transform} in practice, we use two simple results:
\begin{equation}
  \label{eq:hilbert-property-1}
  \frac{\zeta+u}{\zeta(\zeta-u)} = \frac{2}{\zeta-u}-\frac{1}{\zeta}
\end{equation}
and
\begin{equation}
  \label{eq:hilbert-property-2}
  2\mathcal{P}\oint_{C} = \lim_{\epsilon\rightarrow 0}
  \left(
    \oint_{C_{+\epsilon}} + \oint_{C_{-\epsilon}}
  \right),
\end{equation}
where $C_{\pm\epsilon}$ denotes a contour centered at the origin with
radius $|u|\pm \epsilon$. The first essential property of the Hilbert
transform for us is
\begin{equation}
  \label{eq:hilbert_property}
  \mathcal{H}u^{n} = i\, \text{sgn}(n)u^{n},
\end{equation}
such that
\begin{equation}
  \label{eq:hilbert-power}
  \hilbert[u] u^{n} = - n\, \text{sgn}(n) u^{n} = -|n|u^{n}.
\end{equation}
Given \eqref{eq:hilbert_property}, we have
\begin{equation}
  \label{eq:hilbert-J}
  \hilbert \tilde{J}(u) = -\sum_{n}|n| a_{n}  u^{-n}. 
\end{equation}
This gives the first part of \eqref{eq:zero-mode-cylinder}
\begin{align}
  \label{eq:JHJ-first-term}
  (\tilde{J}_{-}(\hilbert \tilde{J})_{+})_{0} = \sum_{n >
  0}[a_{-n}(|n|a_{n}) + (|-n|a_{-n})a_{n}] = 2\sum_{n>0}|n|a_{-n}a_{n}.
\end{align}
The OPE part of the analytic ordering requires an adaptation of its
derivation, starting from \eqref{eq:cylinder-to-plane-ordering}. The
structure of the integral is
\begin{align}
  \label{eq:68}
  \cauchy[u]{v} g(u,v) \tilde{J}\overbracket{(u) \hilbert[v] \tilde{J}}(v)
  =
  \cauchy[u]{v} g(u,v) \hilbert[v]
  \left(
  \tilde{J}\overbracket{(u)\tilde{J}}(v)
  \right)
\end{align}
where
\begin{equation}
  \label{eq:5}
  g(u,v) = \frac{1}{u}\sum_{n=1}^{\infty}\psi_{n}(0)
    \left(\frac{u-v}{v}\right)^{n-1}.
\end{equation}
Then
\begin{align}
  \label{eq:61}
  \hilbert[v](\tilde{J}\overbracket{(u)\tilde{J}}(x))
  &= \hilbert[v] 
    \left[
    \frac{uv/2}{(u-v)^{2}}                              
    \right]\nonumber\\[5pt]
  &= \frac{iu}{2}\mathcal{H}_{v} 
    \left[
    \frac{v(u+v)}{(u-v)^{3}}
    \right]\nonumber\\[5pt]
  &= - \frac{uv(u+v)}{2(u-v)^{3}}\,\text{Sgn}(|u|-|v|)\nonumber\\[5pt]
  &= -(\tilde{J}\overbracket{(u)v\del[v]\tilde{J}}(v)) \,\text{Sgn}(|u|-|v|),
\end{align}
where in the last line we used
\begin{equation}
  \label{eq:69}
  \tilde{J}\overbracket{(u)v\del[v]\tilde{J}}(v) =
  uv
  \left(
  \frac{1/2}{(u-v)^{2}}+ \frac{v}{(u-v)^{3}}
\right) = \frac{uv(u+v)}{2(u-v)^{3}}.
\end{equation}
The third line in \eqref{eq:61} is a result of the following crucial
property of the Hilbert transform. Let $f$ be a function with
decomposition $f= f_{+} + f_{-}$, where $f_{+}$ is analytic inside a
disc $D$ and $f_{-}$ is analytic outside $D$. The Hilbert
transform of $f$ is given by
\begin{align}
  \label{eq:65}
  \mathcal{H}f = i(f_{+}-f_{-}),
\end{align}
and, in particular,
\begin{equation}
  \label{eq:66}
  \mathcal{H}f_{\pm} = \pm i f_{\pm}.
\end{equation}
As the domain of analyticity of the OPE depends on the relative radius
of the insertions, we get the result \eqref{eq:61}. Going back to
\eqref{eq:68} and using the decomposition
\begin{align}
  \label{eq:67}
  \oint_{v} = \lim_{\epsilon\rightarrow 0}
  \left(
    \oint_{C_{|v|+\epsilon}} - \oint_{C_{|v|-\epsilon}}
  \right),
\end{align}
we get
\begin{align*}
  \label{eq:37}
  &- \cauchy[u]{v} \frac{1}{u}
    \sum_{n=1}^{\infty}\psi_{n}(0)\left(\frac{u-v}{v}\right)^{n-1}\tilde{J}\overbracket{(u)v\del[v]\tilde{J}}(v)\,\text{Sgn}(|u|-|v|)\nonumber\\[5pt]
  &=-\sum_{k=2}^{3}\psi_{k}(0)\{Jv\del[v]\tilde{J}\}_{k,m}\nonumber\\[5pt]
  &=-\delta_{m,0}\left(-\frac{1}{12}\times
    \frac{1}{2}+\frac{1}{24}\right) = 0
\end{align*}
Therefore,
\begin{equation}
  \label{eq:70}
  (\tilde{J}(u)\hilbert[v]\tilde{J}(v))_{0} = 2\sum_{n>0}|n|a_{-n}a_{n},
\end{equation}
and the final result for $Q_{2}$ is
\begin{equation}
  \label{eq:71}
  Q_{2} = \sum_{n\neq0}L_{-n}a_{n} + 2iQ\sum_{n>0}|n|a_{-n}a_{n} + \frac{1}{3}\sum_{\substack{n+m+p=0}} :a_{n}a_{m}a_{p}:+
  a_{0}\left(L_{0}-\frac{c+1}{24}\right). 
\end{equation}
One can also easily adapt the generalized Wick theorem for other OPEs and obtain higher charges. The coefficients of the qBO$_{2}$ charges are fixed by explicitly computing the commutators of distinct charges. This computation also reduces to calculating OPEs of composite operators.

\subsection{Derivation of $Q_3$}
\label{sec:derivation-q_3-1}

Here we briefly quote the regularized terms of the charge $Q_{3}$. The
associated current is
\begin{gather}
  \label{eq:72}
  \mathcal{Q}_{4} =
  T^{2} + 6TJ^{2} + 6iQ(T\hilbert[v]J + J^{2}\hilbert[v]J)-
  6Q^{2}(\hilbert[v]J)^{2} +(1+Q^{2})J_{x}^{2} + J^{4}.
\end{gather}
The zero mode of each term is given by
\begin{align}
  \label{eq:73}
  (T^{2})_{0} =2\sum_{n>0}L_{-n}L_{n} + L_{0}^{2}
  -\frac{c+2}{12}L_{0} + \frac{5c^{2}+22c}{2880},
\end{align}
\begin{align}
  \label{eq:74}
  (TJ^{2})_{0} = \sum_{m\neq 0}\sum_{n+p=m} L_{-m}a_{n}a_{p}+ 2
  \left(
  L_{0}-\frac{c}{24}
  \right)\sum_{n>0}a_{-n}a_{n} + 
  \left(
  a_{0}^{2} - \frac{1}{24}
  \right)
  \left(
  L_{0} - \frac{c}{24}
  \right),
\end{align}
\begin{align}
  \label{eq:75}
  (T\hilbert[x] J)_{0} = \sum_{n\neq 0} |n|L_{n}a_{-n},
\end{align}
\begin{align}
  \label{eq:76}
  (J^{2}\hilbert[x]J)_{0}
  = \sum_{m+n+p=0}|p|:a_{m}a_{n}a_{p}: ,
\end{align}
\begin{align}
  \label{eq:77}
  (\hilbert[x]J \hilbert[x]J)_{0} = 2 \sum_{n>0} n^{2}a_{-n}a_{n}
  +\frac{1}{240},
\end{align}
\begin{align}
  \label{eq:78}
  (J_{x}J_{x})_{0} = -(u\del[u]J u\del[u]J)_{0}
  = 2\sum_{n>0}n^{2}a_{-n}a_{n}+\frac{1}{240},
\end{align}
\begin{align}
  \label{eq:79}
  (J^{4})_{0} =(J(J^{3}))_{0}
  &= \sum_{m+n+p+q=0}:a_{m}a_{n}a_{p}a_{q}: -\frac{6}{24}
    \left(
    2\sum_{n>0}a_{-n}a_{n}+a_{0}^{2}
    \right) + \frac{15}{2880}.
\end{align}
The results for the terms only involving $J's$ can be also obtained
from zeta function regularization. The sum of all these terms, after
simplifications, give the charge
\begin{align*}
  Q_{3}
  =& 2\sum_{n>0}L_{-n}L_{n} + L_{0}^{2}
     -\frac{c+5}{12}L_{0} + \frac{5c^{2}+42c+37}{2880}+
     6\sum_{m\neq 0}\sum_{n+p=m} L_{-m}a_{n}a_{p}+ \\[5pt]
   &6\left( L_{0}-\frac{c+1}{24} \right)
     \left(
     2\sum_{n>0}a_{-n}a_{n}
     + a_{0}^{2} 
     \right)
     +6iQ\sum_{n\neq 0} |n|L_{n}a_{-n}+\\[5pt]
   &6iQ\sum_{m+n+p=0}|p|:a_{m}a_{n}a_{p}: + 2(1-5Q^{2})
     \sum_{n>0}n^{2}a_{-n}a_{n}+ \sum_{m+n+p+q=0}:a_{m}a_{n}a_{p}a_{q}:.
\end{align*}
This result slightly differs from \cite{Alba:2010qc} in the vacuum
eigenvalue, if we set Virasoro and U(1) zero modes to zero. We can
trace back this difference to the constant terms in \eqref{eq:77} and
\eqref{eq:78}.

\section{Combinatorial Identities of Partitions}
\label{sec:comb-ident}

First, we show the identity \eqref{eq:comb-ident-1}
\begin{align*}
  \label{eq:24}
  \sum_{i}(2i-1)\lambda_{i} &= \sum_{(i,j)\in \lambda}(2i-1)\\[5pt]
                            &= \sum_{j}\sum_{i=1}^{\lambda_{j}'}(2i-1)\\[5pt]
                            &= \sum_{j}(\lambda_{j}'(\lambda_{j}'+1) -\lambda_{j}')\\[5pt]
  &= \sum_{j}\lambda_{j}'^{2},
\end{align*}
where we used that $\sum_{j=1}^{\lambda_{i}}1 = \lambda_{i}$ and
$\sum_{i=1}^{\lambda_{j}'}i = \frac{\lambda_{j}'(\lambda_{j}'+1)}{2}$
above. The second identity to be shown is \eqref{eq:comb-ident-2}
\begin{align*} 
  \sum_{i}(2i-1)^{2}\lambda_{i}
  &=\sum_{i}(4i^{2}-4i+1)\lambda_{i}\\[5pt]
  &=4\sum_{i}i^{2}\lambda_{i}
    -4\sum_{i}i\lambda_{i}
    +|\lambda|\\[5pt]
  &=4\sum_{(i,j)\in\lambda} i^{2} -2
    \sum_{j}\lambda_{j}'^{2}-|\lambda|\\[5pt]
  &=4\sum_{j}\frac{1}{6}\lambda'_{j}(\lambda'_{j}+1)(2\lambda'_{j}+1) -2 \sum_{j}\lambda_{j}'^{2}-|\lambda|\\[5pt]
  &=\frac{4}{3}\sum_{j}\lambda_{j}'^{3}
    - \frac{1}{3}|\lambda|.
\end{align*}
We can similarly derive the transformed expression for
$\sum_{i}(2i-1)^{p}\lambda_{i} \propto
\sum_{j}{\lambda'_{j}}^{p+1}+\dots$ for all $p\in \mathbb{N}$. These
identities are useful to express parts of the qBO$_{2}$ eigenvalues in
terms of conjugate partitions.

The free boson representation is more useful to integrate the
partition function, for example
\begin{equation}
  \label{eq:7}
  \sum_{i}\lambda_{i}^{k} = \sum_{r} N_{r} r^{k},
\end{equation}
where $N_{r}$ is the number of rows of the partition $\lambda$ with
$r$ boxes. In this representation, partitions are denoted by $\lambda =
(1^{N_{1}}2^{N_{2}}\cdots)$. The challenge here is to obtain the free boson
representation for $\sum_{i}(2i-1)^{q}\lambda_{i}$. Let us see what
happens for $q=1$. Let $n_{i}\equiv N_{\lambda_{i}}$ be the number of rows with
$\lambda_{i}$ number of boxes, in such a way that
\begin{align}
  \label{eq:25}
  \sum_{i}(2i-1)\lambda_{i} &= \sum_{i=1}^{n_{1}} (2i-1)
  \lambda_{1}+\sum_{i=n_{1}+1}^{n_{1}+n_{2}} (2i-1)
  \lambda_{2}+\sum_{i=n_{1}+n_{2}+1}^{n_{1}+n_{2}+n_{3}}
  (2i-1) \lambda_{3}+\cdots\nonumber\\[5pt]
  &=\sum_{k} c_{k}\lambda_{k},
\end{align}
where the $\lambda_{k}$ degeneracy is given by
\begin{equation}
  \label{eq:27}
   c_{k} = \sum^{\sum_{j=1}^{k}n_{j}}_{i=\sum_{j=1}^{k-1}n_{j}+1} (2i-1)
\end{equation}
Now, let
\begin{equation}
  \label{eq:26}
   \tilde{n}_{k} =
          \sum_{j=1}^{k}n_{j},
\end{equation}
then
\begin{align*}
 \sum^{\tilde{n}_{k-1}+n_{k}}_{i=\tilde{n}_{k-1}+1} i       &=
          \sum^{\tilde{n}_{k-1}+n_{k}}_{i=1} i
          -\sum^{\tilde{n}_{k-1}}_{i=1} i\\[5pt]
        &= \frac{(\tilde{n}_{k-1}+n_{k})(\tilde{n}_{k-1}+n_{k}+1)}{2} -
          \frac{\tilde{n}_{k-1}(\tilde{n}_{k-1}+1)}{2}\\[5pt]
        &=\frac{1}{2} (n_{k}^{2} + 2 \tilde{n}_{k-1}n_{k}+n_{k} )\\[5pt]
        &=\frac{1}{2}n_{k}(n_{k}+1) +\tilde{n}_{k-1}n_{k}
\end{align*}
and thus
\begin{equation}
  \label{eq:28}
  c_{k} = n_{k}(n_{k}+1) +2\tilde{n}_{k-1}n_{k} -n_{k} =n_{k}^{2} +2\tilde{n}_{k-1}n_{k}.
\end{equation}
Therefore
\begin{align*}
  \sum_{k}c_{k}\lambda_{k} = \sum_{k}(n_{k}^{2}
  +2\tilde{n}_{k-1}n_{k})\lambda_{k}.
\end{align*}
First, using the definition of $n_{k} = N_{\lambda_{k}}$,  we notice that
\begin{equation}
  \label{eq:29}
  \sum_{k} n_{k}^{2}\lambda_{k} = \sum_{k}
  N_{\lambda_{k}}^{2}\lambda_{k} =\sum_{r} N_{r}^{2}r.
\end{equation}
Second, noticing that $\tilde{n}_{0}=0$,
\begin{align*}
  \sum_{k}n_{k}\tilde{n}_{k-1}\lambda_{k}
  &=
    \sum_{k}N_{\lambda_{k}}\lambda_{k}\sum_{j=1}^{k-1}N_{\lambda_{j}}\\[5pt]
  &=N_{\lambda_{2}}\lambda_{2}N_{\lambda_{1}}+N_{\lambda_{3}}\lambda_{3}(N_{\lambda_{1}}+N_{\lambda_{2}})+N_{\lambda_{4}}\lambda_{4}(N_{\lambda_{1}}+N_{\lambda_{2}}+N_{\lambda_{3}})+\dots\\[5pt]
  &=N_{\lambda_{1}}(N_{\lambda_{2}}\lambda_{2}+N_{\lambda_{3}}\lambda_{3}+\dots)
    +
    N_{\lambda_{2}}(N_{\lambda_{3}}\lambda_{3}+N_{\lambda_{4}}\lambda_{4}+\dots)+N_{\lambda_{3}}(N_{\lambda_{4}}\lambda_{4}+\dots).\\[5pt]
  &=\sum_{i}N_{\lambda_{i}}\sum_{j>i}N_{\lambda_{j}}\lambda_{j}\\
  &=\sum_{r}N_{r}\sum_{s=1}^{r-1}N_{s}s.
\end{align*}
Finally, using all the results above, we get
\begin{equation}
  \label{eq:30}
  \sum_{i}(2i-1)\lambda_{i} = \sum_{r} 
  \left(
    N_{r}^{2}r + N_{r}\sum_{s=1}^{r-1}N_{s}s
  \right) = \sum_{r}N_{r}\sum_{s=1}^{r}N_{s}s.
\end{equation}
Notice that this can be rewritten as
\begin{equation}
  \label{eq:conjugate-partition-identity-k2-appendix}
  \sum_{r}N_{r}\sum_{s=1}^{r}N_{s}s = \sum_{r,s}N_{r}A_{rs}N_{s},
\end{equation}
where $A_{rs}$ is the (infinite) lower triangular matrix
\begin{equation}
  \label{eq:lower-triangular-matrix}
  A=\begin{pmatrix}
    1& 0 & 0&0&\dots \\
    1&2 & 0&0&\dots \\
    1&2 & 3&0 &\dots\\
    \hdotsfor{5}
  \end{pmatrix}.
\end{equation}
We can generalize this derivation to terms like $\sum_{i}(2i-1)\lambda_{i}^{q},
q>1$ giving
\begin{equation}
  \label{eq:96}
  \sum_{r,s}A_{r,s}^{(q)}N_{r}N_{s},\quad
A^{(q)} =  \begin{pmatrix}
        1& 0 & 0&0&\dots \\
    1&2^{q} & 0&0&\dots \\
    1&2^{q} & 3^{q}&0 &\dots\\
    \hdotsfor{5}
  \end{pmatrix}.
\end{equation}
This derivation also suggests that the terms
$\sum_{i}(2i-1)^{p}\lambda_{i},\; p>1$ will be higher-order
polynomials in $N_{r}$
\begin{equation}
  \label{eq:33}
  A^{(p,1)}_{r_{1} r_{2}\dots r_{p+1}}N_{r_{1}}N_{r_{2}}\cdots
  N_{r_{p+1}},\quad r_{1} \geq r_{2} \geq \dots \geq r_{p+1} >0.
\end{equation}
This complicates the integration of these terms in the partition
function, which might be related to the theory of integral
discriminants \cite{morozov2009introduction}. Furthermore, generic
terms like $\sum_{i}(2i-1)^{p}\lambda^{q}_{i},\; p,q>1$ will be just
like \eqref{eq:33}, but the tensors $A^{(p,q)}$ will have components
which are $q$-th powers of integers.

\bibliographystyle{JHEP}
\bibliography{thermal_bo2}

\end{document}